\documentclass{aa}  

\usepackage{graphicx}
\usepackage{txfonts,textcomp}
\usepackage{gensymb} 
\usepackage{xspace}
\usepackage{tabularx}
\usepackage{xcolor}
\usepackage{placeins}
\usepackage{threeparttable}
\usepackage{ulem}
\usepackage{natbib,twoopt}
\usepackage[breaklinks=true]{hyperref} 
\bibpunct{(}{)}{;}{a}{}{,} 
\makeatletter
\newcommandtwoopt{\citeads}[3][][]{\href{http://adsabs.harvard.edu/abs/#3}%
{\def\hyper@linkstart##1##2{}%
\let\hyper@linkend\@empty\citealp[#1][#2]{#3}}}
\newcommandtwoopt{\citepads}[3][][]{\href{http://adsabs.harvard.edu/abs/#3}%
{\def\hyper@linkstart##1##2{}
\let\hyper@linkend\@empty\citep[#1][#2]{#3}}}
\newcommandtwoopt{\citetads}[3][][]{\href{http://adsabs.harvard.edu/abs/#3}%
{\def\hyper@linkstart##1##2{}
\let\hyper@linkend\@empty\citet[#1][#2]{#3}}}
\newcommandtwoopt{\citeyearads}[3][][]%
{\href{http://adsabs.harvard.edu/abs/#3}
{\def\hyper@linkstart##1##2{}%
\let\hyper@linkend\@empty\citeyear[#1][#2]{#3}}}
\makeatother
\def\ms{\hbox{m\,s$^{-1}$}}         
\def\m2s2{\hbox{\,m$^{2}$\,s$^{-2}$}} 
\def\Msun{$M_{\odot}$\xspace}             

\def\Mjup{\hbox{$\mathrm{M}_{\rm J}$}\xspace}
\def\Rjup{\hbox{$\mathrm{R}_{\rm J}$}\xspace}

\def\ten[#1]{$\;\times 10^{#1}$}

\def\logg{$\log g$}

\newcommand{\e}[1]{{\times10^{#1}}}

\newcommand{\Rnom}{\hbox{$\mathcal{R}^{\rm N}_{\odot}$}} 

\newcommand{\GMnom}{\hbox{$\mathcal{(GM)}^{\rm N}_{\odot}$}}

\newcommand{\Renom}{\hbox{$\mathcal{R}^{\rm N}_{e \rm E}$}}

\newcommand{\GMenom}{\hbox{$\mathcal{(GM)}^{\rm N}_{\rm E}$}}

\newcommand{\RJnom}{\hbox{$\mathcal{R}^{\rm N}_{e \rm J}$}}

\newcommand{\GMJnom}{\hbox{$\mathcal{(GM)}^{\rm N}_{\rm J}$}}

\newcommand{\rebound}{{\sc \tt REBOUND}\xspace}
\newcommand{\whf}{{\sc \tt WHFast}\xspace}
\newcommand{\emcee}{{\sc \tt emcee}\xspace}
\newcommand{\juliet}{{\sc \tt juliet}\xspace}
\newcommand{\batman}{{\sc \tt batman}\xspace}

\newcommand{\celerite}{{\sc \tt celerite}\xspace}

\newcommand{\astropy}{{\sc \tt astropy}\xspace}
\newcommand{\prose}{{\sc \tt prose}\xspace}
\newcommand{\photutils}{{\sc \tt photutils}\xspace}

\newcommand{\Lightkurve}{{\sc \tt Lightkurve}\xspace}
\newcommand{\ttvfast}{{\sc \tt TTVFast}\xspace}
\newcommand{\nautilus}{{\sc \tt NAUTILUS}\xspace}

\newcommand{\MEarth}{$\mathrm{M_E}$\xspace}

\newcommand{\red}{\color{red}}
\newcommand{\cyan}{\color{cyan}}
\newcommand{\blue}{\color{blue}}

\newcommand{\be}{\begin{equation}}
\newcommand{\ee}{\end{equation}}
\newcommand{\hand}{\hspace{0.5cm}{\rm and}\hspace{0.5cm}}
\newcommand{\rn}[1]{(\ref{#1})}
\newcommand{\mv}{\mathrm{v}}
\newcommand{\bea}{\begin{eqnarray}}
\newcommand{\eea}{\end{eqnarray}}
\newcommand{\ff}[2]{{\textstyle \frac{#1}{#2}}}
\newcommand{\nextt}{\nonumber \\}
\newcommand{\ei}{e_{\rm i}}
\newcommand{\eo}{e_{\rm o}}
\newcommand{\mi}{m_{\rm i}}
\newcommand{\mo}{m_{\rm o}}
\newcommand{\Pei}{P_{\rm i}}
\newcommand{\Peo}{P_{\rm o}}

\newcommand{\nuo}{\nu_{\rm o}}
\newcommand{\tref}{t_{\rm ref}}
\newcommand{\ttheta}{\tilde\theta}
\newcommand{\omb}{\overline m_{\rm i}}
\newcommand{\omc}{\overline m_{\rm o}}
\newcommand{\ben}{\begin{enumerate}}
\newcommand{\een}{\end{enumerate}}

\def\logg{$\log g$}
\def\Msun{$M_{\odot}$\xspace}            

\newcommand{\orcid}[1]{\protect\href{https://orcid.org/#1}{\protect\includegraphics[width=8pt]{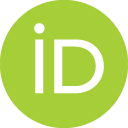}}}
\hypersetup{colorlinks=true, citecolor=blue, linkcolor=blue}

\begin{document}

   \title{Photodynamical modeling of TOI-4504 reveals its deeply resonant state and similarity to GJ~876}

   \author{
        J.M.~Almenara\orcid{0000-0003-3208-9815}\inst{\ref{Geneva}}
        \and R.~Mardling\orcid{0000-0001-7362-3311}\inst{\ref{Monash},\ref{Geneva}} 
        \and A.~Leleu\orcid{0000-0003-2051-7974}\inst{\ref{Geneva}} 
        \and S.~Udry\orcid{0000-0001-7576-6236}\inst{\ref{Geneva}}    
        \and T.~Forveille\orcid{0000-0003-0536-4607}\inst{\ref{Grenoble}}
        \and X.~Bonfils\orcid{0000-0001-9003-8894}\inst{\ref{Grenoble}}
        \and F.~Bouchy\orcid{0000-0002-7613-393X}\inst{\ref{Geneva}} 
        \and C.~Cadieux\orcid{0000-0001-9291-5555}\inst{\ref{Geneva}}
        \and J.~Couturier\inst{\ref{Geneva}}
        \and R.F.~D\'{i}az\orcid{0000-0001-9289-5160}\inst{\ref{ITBA},\ref{BA}}
        \and Y.~Eyholzer\inst{\ref{Geneva}}
        \and E.~Fontanet\orcid{0000-0002-0215-4551}\inst{\ref{Geneva}}
        \and T.~Guillot\orcid{0000-0002-7188-8428}\inst{\ref{Nice}}
        \and G.~H\'ebrard\inst{\ref{iap},\ref{ohp}}  
        \and R.M.~Hoogenboom\orcid{0009-0004-9519-9143}\inst{\ref{Geneva}} 
        \and J.~Korth\orcid{0000-0002-0076-6239}\inst{\ref{Geneva}}
        \and M.~Lendl\orcid{0000-0001-9699-1459}\inst{\ref{Geneva}} 
        \and A.~Nigioni\orcid{0009-0004-5882-6574}\inst{\ref{Geneva}}
        }
   \institute{
        Observatoire de Gen\`eve, Département d’Astronomie, Universit\'e de Gen\`eve, Chemin Pegasi 51b, 1290 Versoix, Switzerland\label{Geneva}
        \and School of Physics and Astronomy, Monash University, Victoria, 3800, Australia\label{Monash}
        \and Univ. Grenoble Alpes, CNRS, IPAG, F-38000 Grenoble, France\label{Grenoble}
        \and Instituto Tecnol\'ogico de Buenos Aires (ITBA), Iguaz\'u 341, Buenos Aires, CABA C1437, Argentina\label{ITBA}
        \and Instituto de Ciencias F\'isicas (ICIFI; CONICET), ECyT-UNSAM, Campus Miguelete, 25 de Mayo y Francia, (1650) Buenos Aires, Argentina\label{BA}
        \and Universit\'e C\^ote d'Azur, Laboratoire Lagrange, OCA, CNRS UMR 7293, Nice, France\label{Nice}
        \and Institut d'astrophysique de Paris, UMR7095 CNRS, Universit\'e Pierre \& Marie Curie, 98bis boulevard Arago, 75014 Paris, France\label{iap}
        \and Observatoire de Haute-Provence, 04670 Saint Michel l'Observatoire, France\label{ohp}
             }

   \date{}

 
  \abstract
   {
   The K-dwarf TOI-4504 hosts two giant planets in 2:1 mean-motion resonance, with orbital periods of 41.3 days (planet~d) and 82.8~days (planet~c). They exhibit among the largest known absolute transit-timing variations, with respective peak-to-node amplitudes up to 5 and 3~days. Newer TESS data show that the previously non-transiting planet~d has now precessed into transiting, and we derive updated system parameters with significant discrepancies with the discovery paper.
The revised parameters place planets~d and c deep in the resonance and
 close to or in the fully-relaxed limit-cycle state, with the resonant and secular modes
 interfering nonlinearly to induce non-zero relaxed free eccentricities which precess at the same
 rate as the forced eccentricities and the longitude of conjunctions, 
 in turn enabling precise measurement of the full
 eccentricities and apsidal angles. 
 We discuss the predictions of linear theory and how it can be used to understand the true state of the system
 revealed by $N$-body integrations, and more generally why it is that the posteriors of
systems more compact than 2:1 tend to suffer from significant eccentricity degeneracy. We show that the extraordinary 
 dynamical states of the giant pairs orbiting TOI-4504 and the M-dwarf GJ~876 are remarkably similar,
 in spite of the significant difference in their host-star masses, and discuss the implications for 
 damping timescales during the relatively gentle formation process of Type II migration.

   }


   \maketitle

\section{Introduction}

\citet{Vitkova2025} discovered the TOI-4504 system with the Transiting Exoplanet Survey Satellite \citep[TESS;][]{Ricker2015}. They found that the K1V stellar host is transited by a sub-Neptune in a 2.4-days orbit (planet~b) and by a Jupiter-sized planet with a period of 82.5~days (planet~c). TOI-4504\,c exhibited among the largest absolute transit-timing variations \citep[TTVs;][]{Agol2005,Holman2005} known, with a peak-to-node amplitude of $\sim$2~days. The large TTVs indicate a strong dynamical interaction. Using radial-velocity (RV) measurements, \citet{Vitkova2025} identified the perturber, TOI-4504\,d, as a non-transiting  inner giant planet in 2:1 mean-motion resonance (MMR) with planet~c.

TESS obtained additional photometry of TOI-4504 in 2025, which covers three additional transits of TOI-4504\,c and even more importantly reveals that the previously non-transiting planet~d has precessed into a transiting configuration. These observations motivate our reanalysis of the system. We report significantly different system parameters compared to the discovery paper, and investigate the dynamical implications of these updated parameters.

The article is organized as follows: We describe the observations of TOI-4504 in Sect.~\ref{section:observations}. Section~\ref{section:analysis} presents the analysis of the transit data and the literature RV measurements. Finally, we discuss our results in Sect.~\ref{section:results} which includes a comparison with the results of \citet{Vitkova2025}, as well as a detailed examination of the relaxed dynamical state of the giant pair in TOI-4504, its similarities with the giant pair in GJ~876, and the implications for Type-II migration. The formalism used to draw our conclusions is summarized in Appendix~\ref{FOT} which includes a list of systems in various related dynamical states.

\section{Observations}\label{section:observations}

TOI-4504 lies in the TESS southern continuous viewing zone, near the south ecliptic pole, and was observed in 43 sectors between 2018 and 2026. During Sectors 1-13, it was monitored only in full-frame images (FFIs), with a cadence of 1800~seconds. TOI-4504 was later upgraded to a 120~seconds cadence in Sectors 27-38, 61-65, 67-69, 87-90, and 94-98. When available, we used the presearch data-conditioning simple aperture photometry (PDCSAP; \citealt{Smith2012}, \citealt{Stumpe2012,Stumpe2014}) light curves of TOI-4504, produced by the TESS Science Processing Operations Center \citep[SPOC;][]{Jenkins2016,Caldwell2020}, with the exception of the transit of planet~c in Sector 61 which is only available in simple aperture photometry. For the observations in FFIs only, we produced light curves using {\it TESSCut} \citep{Brasseur2019} from FFIs calibrated by SPOC, and performed photometry with the \Lightkurve package \citep{Lightkurve2018}\footnote{BJD$_{\rm TDB}$ times for the target observations were computed following \url{https://github.com/havijw/tess-time-correction}.}. Prior to the availability of the light curve produced by SPOC for Sector 98, we made use of the TESS Image CAlibrator Full Frame Images \citep[TICA;][]{Fausnaugh2020}, which have a cadence of 200~seconds.

We analyzed a dataset consisting of 27-hour segments centered on the transits of planets~c and d, corresponding to approximately three times the transit duration of planet c. For planet~d, we also included the epochs of minimum star-planet projected separation in Sectors 62–69, even though no transits were detected, in order to exclude transiting configurations at these times. The shorter transits of planet~b were incorporated only from sectors with 120-second cadence, using shorter segments of 7.5~hours around each transit. 

In addition to the TESS data, we included ground‑based observations of the egress of planet c on 12 January 2026 and the full transit of planet~d on 22 January 2026, obtained with EulerCam \citep[ECAM;][]{Lendl2012} on the Swiss 1.2~m Euler telescope at La Silla Observatory. Observations were conducted with the NGTS filter, using exposure times ranging from 40 to 80 seconds. The image reduction was carried out following \citet{Lendl2012} and \citet{Lendl2014}. Aperture photometry was performed using \prose \citep{Garcia2022}, which relies on \astropy \citep{astropy2022} and \photutils \citep{Bradley2023}. The optimal differential photometry followed \citet{Broeg2005}.
Alongside these photometric datasets, we also included the RV measurements obtained with the FEROS spectrograph \citep{Kaufer1999} and reported by \citet{Vitkova2025}.

\section{Analysis}\label{section:analysis}

We jointly fit the observed photometry and RVs using a photodynamical model \citep{Carter2011} that accounts for the gravitational interactions among the four known bodies in the system. Their positions and velocities were computed as a function of time using the $N$-body code \rebound \citep{Rein2012} employing the \whf integrator \citep{Rein2015} with a time step of 0.02~days. The sky-projected positions were used to generate the light curve with \batman \citep{Kreidberg2015}, incorporating the light-time effect \citep{Irwin1952}. Oversampling was applied to the TESS data to account for the integration-induced distortion described by \citet{Kipping2010}. The line-of-sight velocity of the star, derived from the $N$-body integration, was used to model the RV measurements. The model was parameterized using the stellar mass and radius, limb-darkening coefficients, planet-to-star mass and radius ratios, and Jacobi orbital elements (Table~\ref{table:results}) at the reference epoch ($t_{\mathrm{ref}}=2\,460\,886.2838$~BJD$_{\mathrm{TDB}}$). Due to the symmetry of the problem, we fixed the longitude of the ascending node of planet~c at $t_{\mathrm{ref}}$ to 180\degree, and constrained the inclination of planet~d to be less than 90\degree. For the ECAM observations, we included a linear model in the peak value of the point-spread function and the airmass, and we modeled the error terms using Gaussian-process regression, adopting the approximate Matern kernel implemented in \celerite \citep{Foreman-Mackey2017}.
The model for the RV has two additional parameters: the systemic velocity and a jitter term. In total, the model comprises 41 free parameters. We adopted normal priors for the stellar mass and radius from \citet{Vitkova2025}, and uninformative priors for the remaining parameters. The joint posterior distribution was sampled using the \emcee algorithm \citep{Goodman2010, emcee}.

\begin{table*}
  \small
\renewcommand{\arraystretch}{1.1}
\setlength{\tabcolsep}{4pt}
\centering
\caption{Inferred system parameters.}\label{table:results}
\begin{tabular}{lccccc}
\hline
Parameter & Units & Prior & Median and 68.3\% CI &  &  \\
\hline
\emph{\bf Star} \\
Stellar mass, $m_\star$              & (\Msun)     & $N(0.89, 0.06)$   & $0.877 \pm 0.055$ \\
Stellar radius, $R_\star$            & (\Rnom)     & $N(0.92, 0.04)$   & $0.931 \pm 0.026$ \\
Stellar mean density, $\rho_{\star}$ & ($\mathrm{g\;cm^{-3}}$) &       & $1.53 \pm 0.11$ \\
Surface gravity, \logg\              & (cgs)       &                   & $4.442 \pm 0.025$ \\
$q_1$ TESS, ECAM                     &             & $U(0, 1)$         & $0.30^{+0.25}_{-0.14}$, $0.23^{+0.27}_{-0.14}$ \\
$q_2$ TESS, ECAM                     &             & $U(0, 1)$         & $0.32^{+0.37}_{-0.21}$, $0.22^{+0.35}_{-0.17}$  \smallskip\\

\emph{\bf Planets} & &  & \emph{{\bf Planet~b}} & \emph{{\bf Planet~d}} & \emph{{\bf Planet~c}} \\
Semi-major axis, $a$                      & (au)                   &                                           & $0.03382 \pm 0.00072$   & $0.2221 \pm 0.0047$          & $0.3595 \pm 0.0076$               \\
Eccentricity, $e$                         &                        &                                           & < 0.74$^\dagger$        & $0.2930 \pm 0.0012$          & $0.04989 \pm 0.00023$             \\
Argument of periastron, $\omega$          & (\degree)              &                                           & $180 \pm 160$           & $195.97 \pm 0.15$            & $178.51 \pm 0.54$                 \\
Inclination, $i$                          & (\degree)              & $S(0, 180)_{\rm b, c}$, $S(0, 90)_{\rm d}$& $92.5^{+2.0}_{-4.2}$    & $88.981 \pm 0.036$           & $89.543 \pm 0.034$                \\
Longitude of the ascending node, $\Omega$ & (\degree)              & $U(90, 270)_{\rm b,d}$                    & $215 \pm 33$            & $182.03 \pm 0.26$            & 180 (fixed at $t_{\mathrm{ref}}$) \\
Mean anomaly, $M_0$                       & (\degree)              &                                           & $196^{+76}_{-41}$       & $166.28 \pm 0.26$            & $277.30 \pm 0.55$                   \\
$\sqrt{e}\cos{\omega}$                    &                        & $U(-1, 1)$                                & $-0.15^{+0.79}_{-0.41}$ & $-0.5205 \pm 0.0013$         & $-0.22328 \pm 0.00049$            \\
$\sqrt{e}\sin{\omega}$                    &                        & $U(-1, 1)$                                & $-0.01 \pm 0.24$        & $-0.1489 \pm 0.0012$         & $0.0058 \pm 0.0021$               \\
Mass ratio, $m_{\mathrm{p}}/m_\star$      &                        & $U(0, 1)$                                 & < $35\e{-5}^\dagger$    & $(220.49 \pm 0.69)\e{-5}$    & $(281.66 \pm 0.34)\e{-5}$         \\
Radius ratio, $R_{\mathrm{p}}/R_\star$    &                        & $U(0, 1)$                                 & $0.02603 \pm 0.00085$   & $0.1103 \pm 0.0033$          & $0.1102 \pm 0.0020$               \\
Scaled semi-major axis, $a/R_{\star}$     &                        &                                           & $7.81 \pm 0.18$         & $51.3 \pm 1.2$               & $83.0 \pm 2.0$                    \\
Impact parameter, $b$                     &                        &                                           & $0.37 \pm 0.25$         & $0.906 \pm 0.013$            & $0.659 \pm 0.036$                 \\
$T_0'$\;-\;2\;460\;000                    & (BJD$_{\mathrm{TDB}}$) & $U(-114, 1887)$                           & $887.1883 \pm 0.0030$   & $900.0759 \pm 0.0017$        & $886.2585 \pm 0.0016$             \\
$P'$                                      & (d)                    & $U(0, 1000)$                              & $2.42634 \pm 0.00014$   & $40.8204 \pm 0.0024$         & $84.0725 \pm 0.0049$              \\
$K'$                                      & (\ms)                  &                                           & < 57$^\dagger$          & $136.5 \pm 2.9$              & $131.2 \pm 2.8$                   \\
Planet mass, $m_{\mathrm{p}}$             &(\MEarth)               &                                           & < 100$^\dagger$         & $644 \pm 40$                 & $822 \pm 51$                      \\
                                          &(\Mjup)                 &                                           & < 0.32$^\dagger$        & $2.03 \pm 0.13$              & $2.59 \pm 0.16$                   \\
Planet radius, $R_{\mathrm{p}}$           &(\Renom)                &                                           & $2.65 \pm 0.12$         & $11.19 \pm 0.54$             & $11.18 \pm 0.43$                  \\
                                          &(\RJnom)                &                                           & $0.236 \pm 0.011$       & $0.999 \pm 0.048$            & $0.998 \pm 0.038$                 \\
Planet mean density, $\rho_{\mathrm{p}}$  &($\mathrm{g\;cm^{-3}}$) &                                           & < 31$^\dagger$          & $2.52 \pm 0.34$              & $3.22 \pm 0.40$                   \\
Planet surface gravity, $\log$\,$g_{\mathrm{p}}$ &(cgs)            &                                           & $3.81^{+0.23}_{-0.40}$  & $3.702 \pm 0.042$            & $3.808 \pm 0.037$                 \\
Equilibrium temperature, T$_{\rm eq}$     & (K)                    &                                           & $1345 \pm 22$           & $525.0 \pm 8.6$              & $412.6 \pm 6.8$                   \\

Mutual inclination, $I_{\rm d,c}$             & (\degree)              &   &  & \multicolumn{2}{c}{$2.11 \pm 0.25$}  \smallskip\\

\emph{\bf RV} \\
FEROS jitter &           &  $U(0, 20)$          & $7.3 \pm 1.0$ \\
FEROS offset & (\ms)     & $U(-10^{5}, 10^{5})$ & $2070 \pm 14$  \smallskip\\

\emph{{\bf Linear ephemerides}} \\
Period                  & (d)                    & & 2.4261518   & 41.2872390  & 82.8190895  \\
$T_0$\;-\;2\;450\;000   & (BJD$_{\mathrm{TDB}}$) & & 8395.529998 & 8423.049769 & 8400.937342 \\

\hline
\end{tabular}
\tablefoot{\tiny The table lists: Prior, posterior median, and 68.3\% credible interval (CI) for the photodynamical analysis (Sect.~\ref{section:analysis}). The Jacobi orbital elements are given for the reference epoch $t_{\mathrm{ref}}=2\,460\,886.2838$~BJD$_{\mathrm{TDB}}$. The parameters $q_1$ and $q_2$ are the quadratic limb-darkening coefficients parameterized using \citet{kipping2013}. $^\dagger$ Upper limit at 95\% confidence. The planetary equilibrium temperature is computed for zero albedo and full day-night heat redistribution. $P'$ and $T_0'$ should not be confused with the linear ephemeris, and they were only used to reduce the correlations between jump parameters, replacing the semi-major axis and the mean anomaly at $t_{\mathrm{ref}}$. The linear ephemerides are derived from the median posterior transit times spanning the years 2018 to 2027. \\$T'_0 \equiv t_{\mathrm{ref}} - \frac{P'}{2\pi}\left(M_0-E+e\sin{E}\right)$ with $E=2\arctan{\left\{\sqrt{\frac{1-e}{1+e}}\tan{\left[\frac{1}{2}\left(\frac{\pi}{2}-\omega\right)\right]}\right\}}$, $P' \equiv \sqrt{\frac{4\pi^2a^{3}}{\mathcal G m_{\star}}}$, $K' \equiv \frac{m_p \sin{i}}{m_\star^{2/3}\sqrt{1-e^2}}\left(\frac{2 \pi \mathcal G}{P'}\right)^{1/3}$. CODATA 2018: $\mathcal G = 6.674\,30$\ten[-11]$\rm{m^3\,kg^{-1}\,s^{-2}}$. IAU 2012: au = $149\,597\,870\,700$~m$\,$. IAU 2015: \Rnom = 6.957\ten[8]~m, \GMnom = 1.327$\,$124$\,$4\ten[20]~$\rm{m^3\,s^{-2}}$, \Renom~=~6.378$\,$1\ten[6]~m, \GMenom = 3.986$\,$004\ten[14]~$\rm{m^3\,s^{-2}}$, \RJnom~=~7.149$\;$2\ten[7]~m, \GMJnom = 1.266$\;$865$\;$3\ten[17]~$\rm{m^3\;s^{-2}}$. \Msun$ = \GMnom/\mathcal G$, \MEarth = \GMenom/$\mathcal G$, \Mjup = \GMJnom/$\mathcal G$, $k^2$ = \GMnom$\,(86\,400~\rm{s})^2$/$\rm{au}^3$. $N(\mu, \sigma)$: Normal distribution with mean $\mu$ and standard deviation $\sigma$. $U(a, b)$: A uniform distribution defined between a lower $a$ and upper $b$ limit. $S(a, b)$: A sinusoidal distribution defined between a lower $a$ and upper $b$ limit.}
\end{table*}

\begin{figure}
  \centering
  \includegraphics[width=0.242\textwidth]{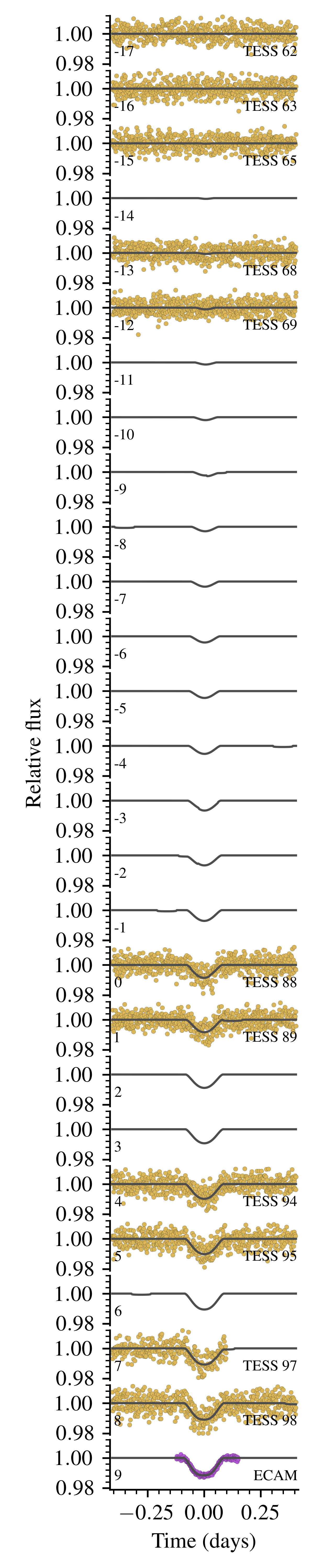}
  \includegraphics[width=0.242\textwidth]{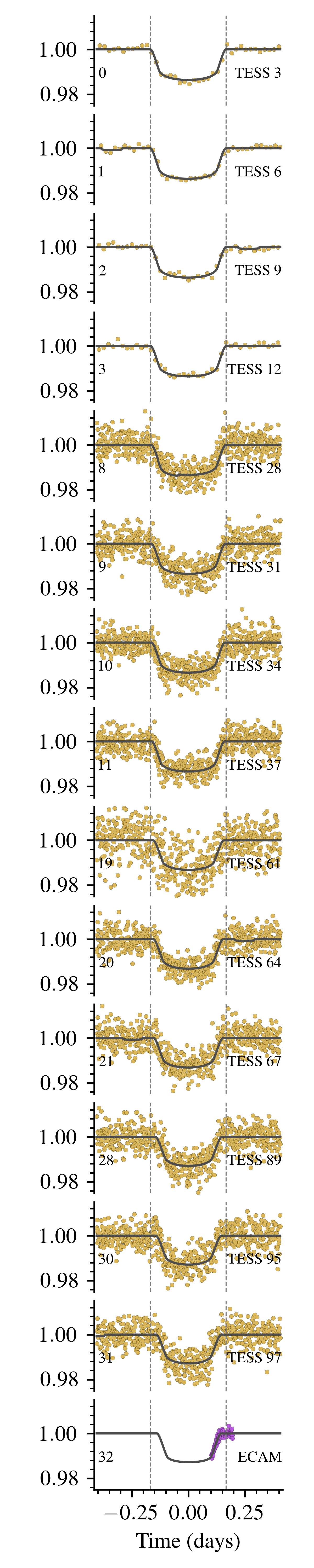}
  \caption{Photodynamical modeling of the transits of planets~d (left) and c (right). The dots represent the observations. The black line shows the oversampled transit model. Each panel is centered at the MAP transit timing and labeled with the epoch number and the sector for TESS observations. For planet~c, the dashed vertical lines delimit a transit duration of 8 hours, close to the maximum value during the observations.} \label{figure:Transits}
\end{figure}

\begin{figure}
  \centering
  \includegraphics[width=0.49\textwidth]{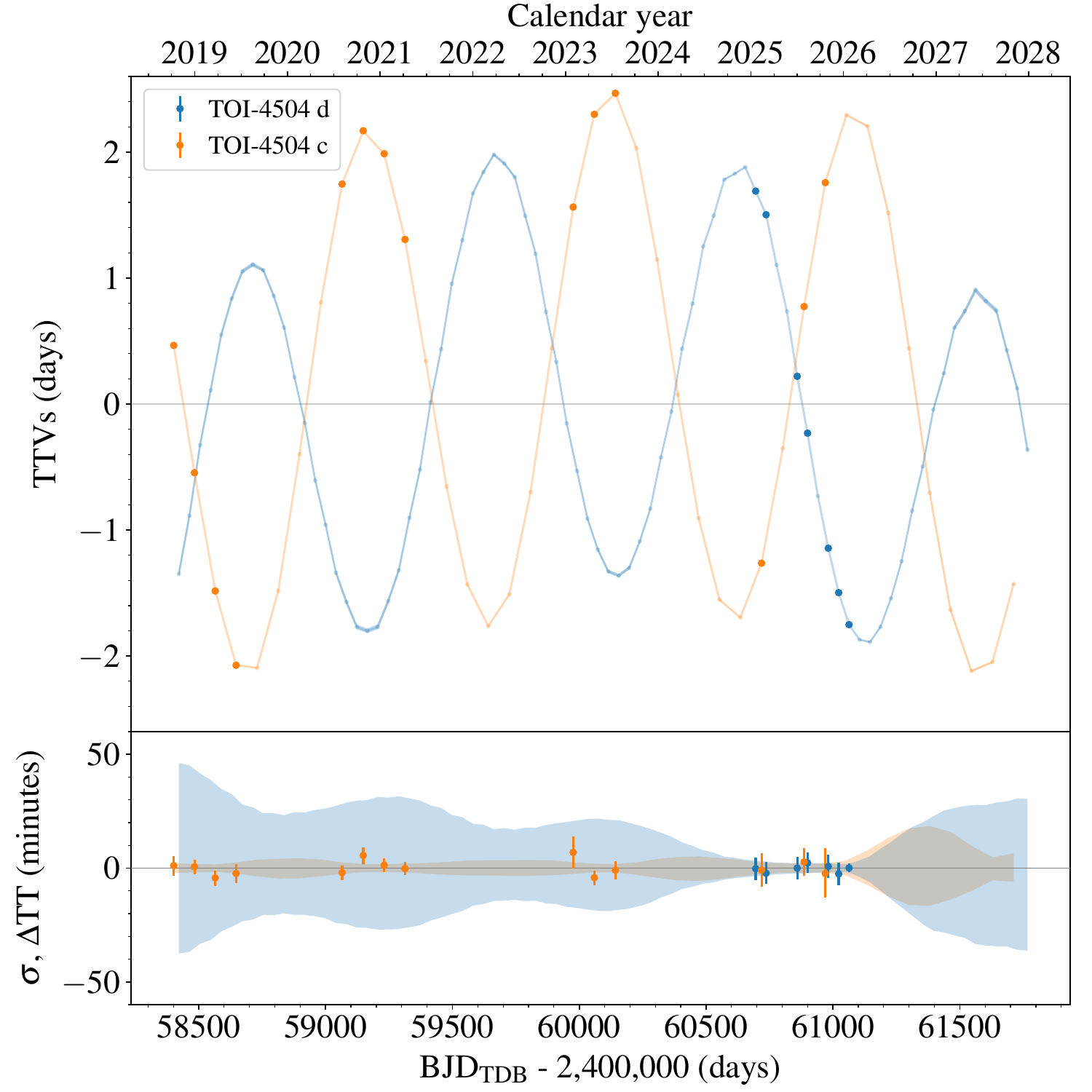}
  \caption{Posterior TTV predictions of planets~d (blue) and c (orange) computed relative to a linear ephemeris (Table~\ref{table:results}). A thousand random draws from the posterior distribution were used to estimate the median TTV values and their uncertainties (68.3\% confidence interval). The upper panel shows the posterior TTV values and compares them with the individual transit-time determinations (Table~\ref{table:transit_times}, error bars). In the lower panel, the posterior median transit-timing value was subtracted to emphasize the uncertainty in the distribution and facilitate a clearer comparison with the individually determined transit times.} \label{figure:TTVs}
\end{figure}

\begin{figure}
  \centering
  \includegraphics[width=0.49\textwidth]{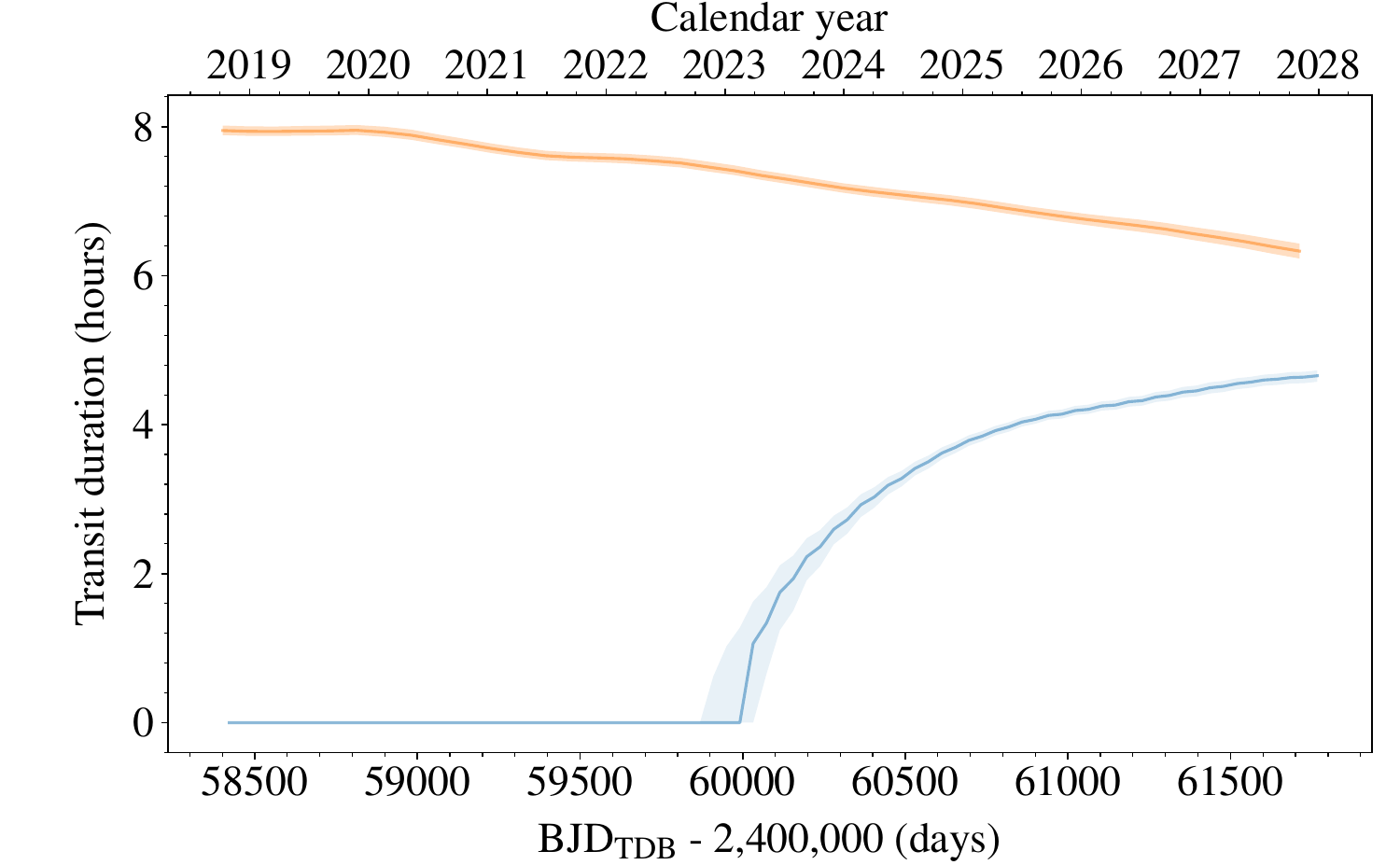}
  \caption{Posterior of the transit duration of planets~d (blue) and c (orange). We used a thousand random draws from the posterior distribution to estimate the median transit duration and its uncertainty (68.3\% CI). A duration of zero correspond with the planet not transiting.} \label{figure:TDVs}
\end{figure}

\begin{figure}
\centering
  \includegraphics[width=0.49\textwidth]{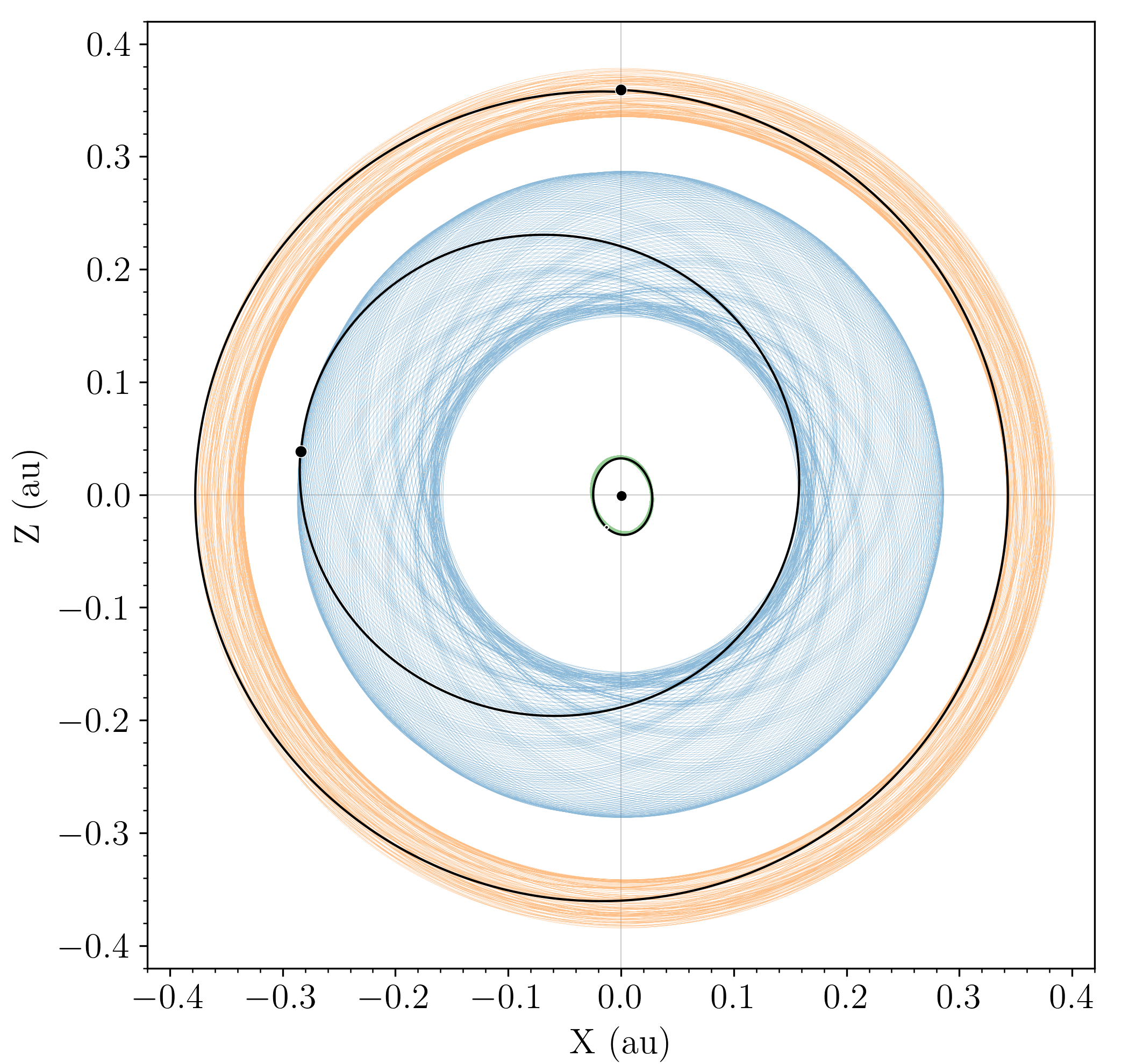}
  \caption{Orbital projection of the MAP model over the time-span of the superperiod for planet~b (green), d (blue), and c (orange). The origin is the system barycenter, and the orbits are projected in the X-Z plane (system top view, movement is clockwise, the positive Z-axis points towards the observer). One orbit near $t_{\mathrm{ref}}$ is shown in black. The black circles mark the position of the star (size to scale) and the planets (enlarged by a factor of 10) at $t_{\mathrm{ref}}$.} \label{figure:orbits}
\end{figure}

\begin{figure}
  \includegraphics[width=0.49\textwidth]{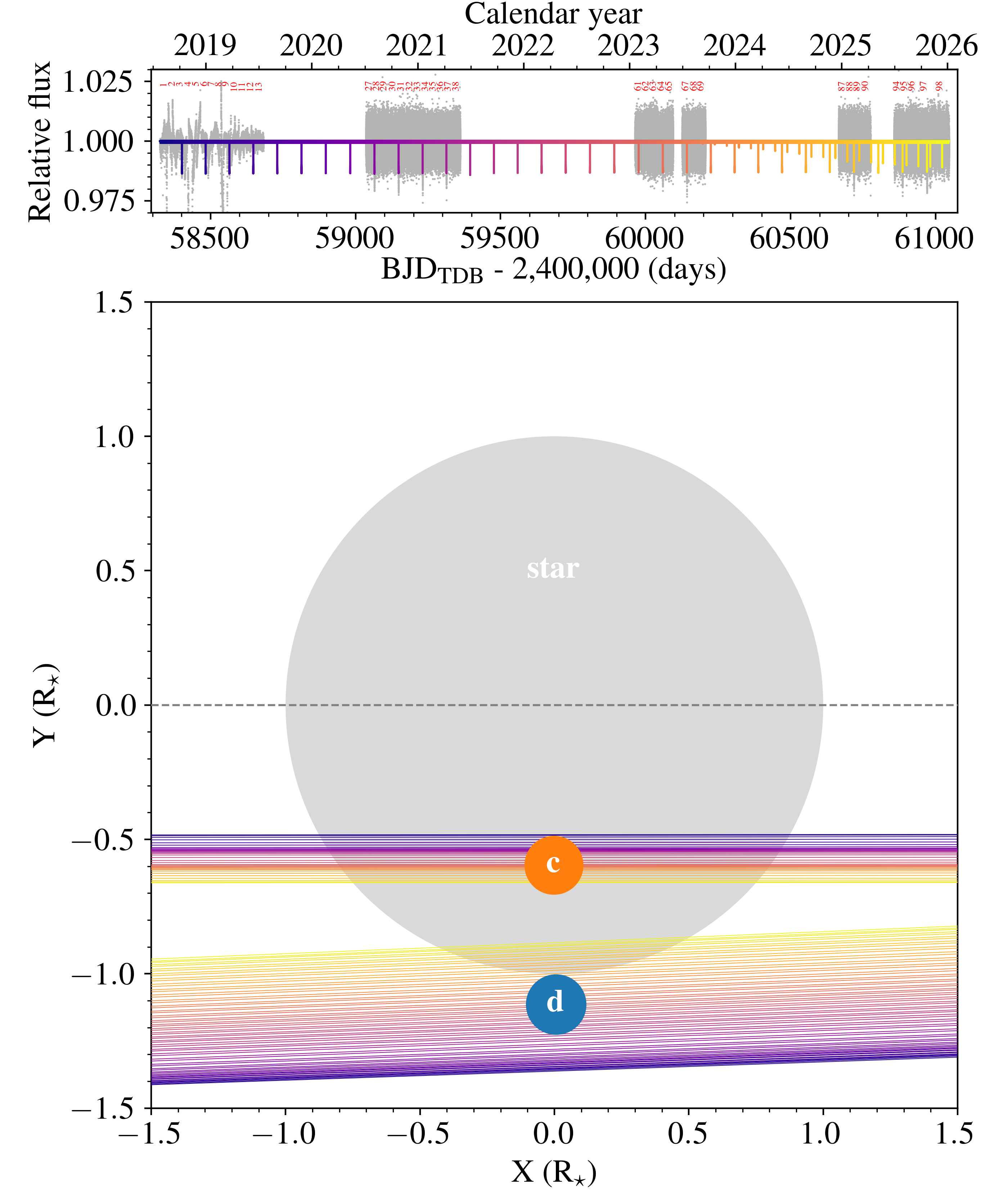}
  \caption{Stellar‑crossing paths of planets~d and~c as seen by the observer for the MAP model. The stellar disc is shown in gray, and the MAP planet radii are indicated by blue (planet~d) and orange (planet~c) disks. The upper panel displays the TESS light curve (gray dots) together with the MAP transit model, with the model color‑coded by time. The same color scale is used in the main panel for the crossing paths. The TESS sector is annotated in red.} \label{figure:transit_path}
\end{figure}

The maximum a posteriori (MAP) model for the transit photometry is plotted in Fig.~\ref{figure:Transits}. Table~\ref{table:results} lists the median and the 68\% credible interval (CI) of the marginal distribution of the inferred system parameters. 
Figure~\ref{figure:TTVs} present the posterior TTVs for planets~d and c, derived from the times of minimum projected separation between the star and planet, and a comparison with the individually determined transit times (Table~\ref{table:transit_times}) computed with \juliet \citep{Espinoza2019,Kreidberg2015,Speagle2020}, assuming a constant transit duration. Figure~\ref{figure:TDVs} shows the transit duration for planets~d and c, calculated from the difference of the first and fourth contact times, as obtained from the sky‑projected star–planet separation. Figures~\ref{figure:orbits} and \ref{figure:transit_path} show, respectively, the posterior top-view planet orbits and the stellar-crossing path as seen by the observer, both derived from the MAP model.

\begin{figure*}
  \centering
  \includegraphics[width=0.99\textwidth]{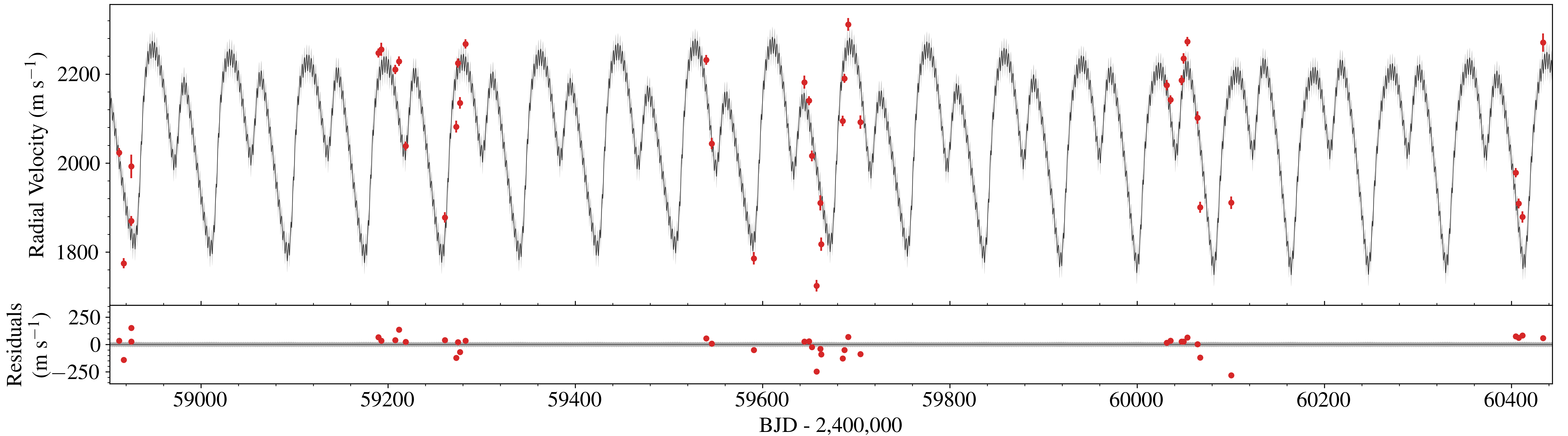}
  \caption{FEROS RVs of TOI-4504 (red error bars) together with the median RV model (black line) and the 68.3\% confidence interval (gray region), both estimated from a thousand random draws from the posterior distribution. Residuals from the median model are shown.} \label{figure:RV}
\end{figure*}

Our model provides a poor fit to the FEROS RVs (Fig.~\ref{figure:RV}), as did \citet{Vitkova2025}'s. The posterior value of the FEROS multiplicative jitter is $7.3\pm1.0$, corresponding to approximately $86$~m/s effective RV uncertainties. We detect no significant periodicity in the residuals that would indicate that the system contains an additional planet. Introducing into a fit of the RV data alone a planet between planets~b and d, beyond~c, or both, does not significantly reduce the jitter. The excess signal may therefore be at least partially attributable to stellar activity. \citet{Vitkova2025} reported that several activity indicators (Bisector, H$\alpha$, He\,\textsc{i}, Na\,\textsc{ii}) show no significant periodicities, but the $\log R'_{\mathrm{HK}}$ index shows multiple marginally significant periods, indicating that TOI‑4504 is an active star. The $\log R'_{\mathrm{HK}}$ index, with weighted mean and standard deviation of $-4.43\pm0.38$, reaches values as high as $-3.92\pm0.05$, which could imply RV scatter of tens to a few hundreds of m/s \citep{Hojjatpanah2020}. It is also possible that the FEROS uncertainties are underestimated. Finally, additional planets remain a possible explanation. 

\begin{figure}
  \centering
  \includegraphics[width=0.48\textwidth]{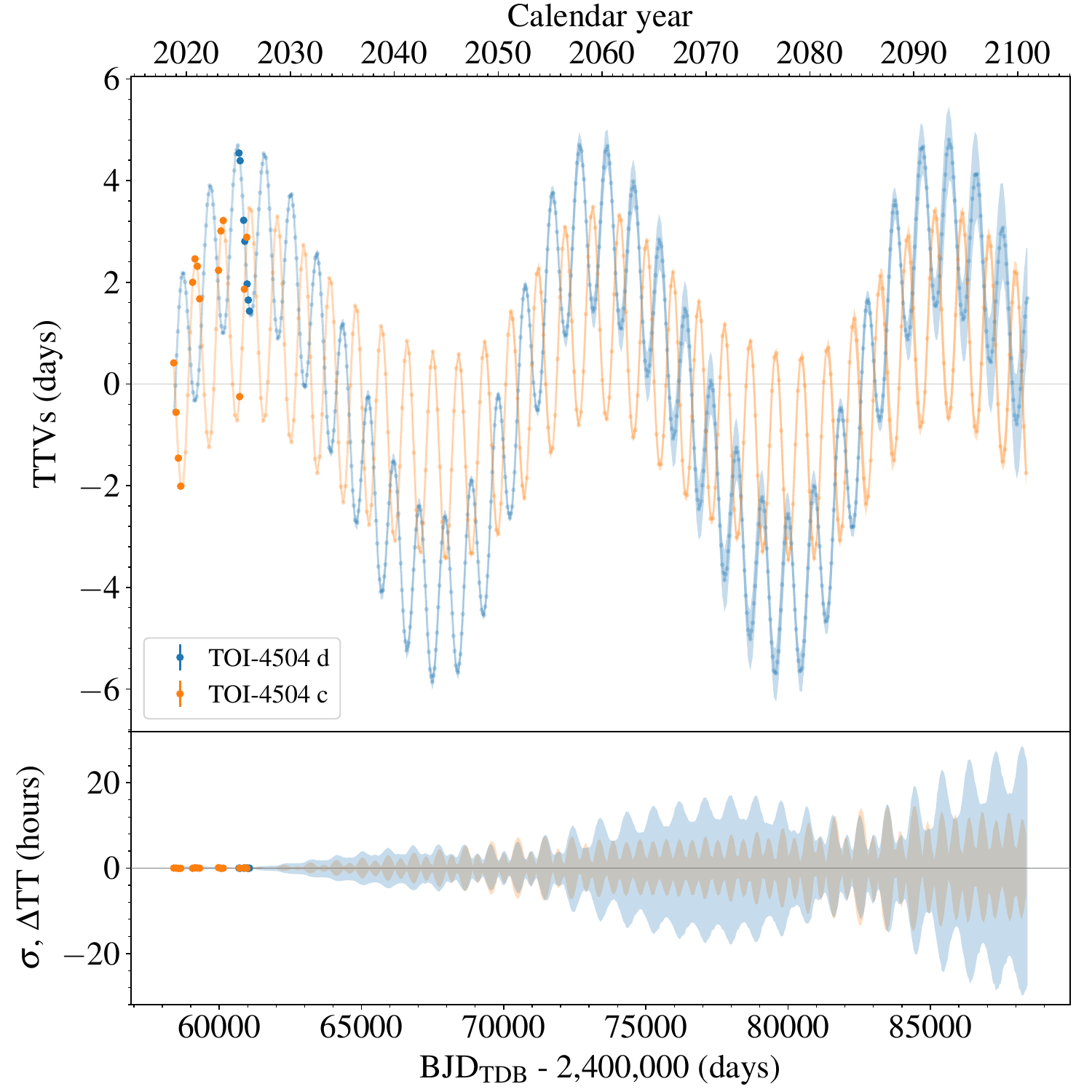}
  \caption{Same as Fig.~\ref{figure:TTVs}, but up to the year 2100. With linear ephemeris: 2458422.235250 + 41.2502075\;$\times$\;epoch (BJD$_{\rm TDB}$), 2458400.988431 + 82.7810389\;$\times$\;epoch (BJD$_{\rm TDB}$) for planet b and c, respectively.} \label{figure:TTVsLongTerm}
\end{figure}

\begin{table}[!b]
\centering
    \small
    \renewcommand{\arraystretch}{1.25}
\caption{Transit times of the observations.}\label{table:transit_times}
\begin{tabular}{rlr}
\hline
Epoch & Posterior median  & TESS Sector / \\
      & and 68.3\% CI [BJD$_{\mathrm{TDB}}$]  & Instrument  \\
\hline
\emph{\bf Planet~d}\\ 
0	& $2460695.5378 \pm 0.0035$          & 88 \\
1	& $2460736.6378_{-0.0033}^{+0.0034}$ & 89 \\
4	& $2460859.2170_{-0.0035}^{+0.0034}$ & 94 \\
5	& $2460900.0533 \pm 0.0031$          & 95 \\
7	& $2460981.7147_{-0.0036}^{+0.0035}$ & 97 \\
8	& $2461022.6487 \pm 0.0034$          & 98 \\
9   & $2461063.6822_{-0.0013}^{+0.0014}$ & ECAM \\

\emph{\bf Planet~c}\\ 
0	& $2458401.4028 \pm 0.0030$          &  3 \\
1	& $2458483.2113 \pm 0.0022$	         &  6 \\
2	& $2458565.0920_{-0.0024}^{+0.0023}$ &  9 \\
3	& $2458647.3219_{-0.0028}^{+0.0027}$ & 12 \\
8	& $2459065.2363 \pm 0.0022$          & 28 \\
9	& $2459148.4782_{-0.0025}^{+0.0024}$ & 31 \\
10	& $2459231.1147 \pm 0.0021$	         & 34 \\
11	& $2459313.2536 \pm 0.0020$          & 37 \\
19	& $2459976.0635_{-0.0050}^{+0.0049}$ & 61 \\
20	& $2460059.6186 \pm 0.0022$          & 64 \\
21	& $2460142.6049 \pm 0.0028$	         & 67 \\
28	& $2460718.6090_{-0.0051}^{+0.0052}$ & 89 \\
30	& $2460886.2837 \pm 0.0043$          & 95 \\
31	& $2460970.0864_{-0.0075}^{+0.0077}$ & 97 \\
32  & $\dagger$                          & ECAM \\
\hline
\end{tabular}
\tablefoot{$\dagger$ Only the egress was observed, and the planet exhibits transit‑duration variations, so an independent transit time cannot be derived.}
\end{table}

\begin{figure}
  \centering
  \includegraphics[width=0.49\textwidth]{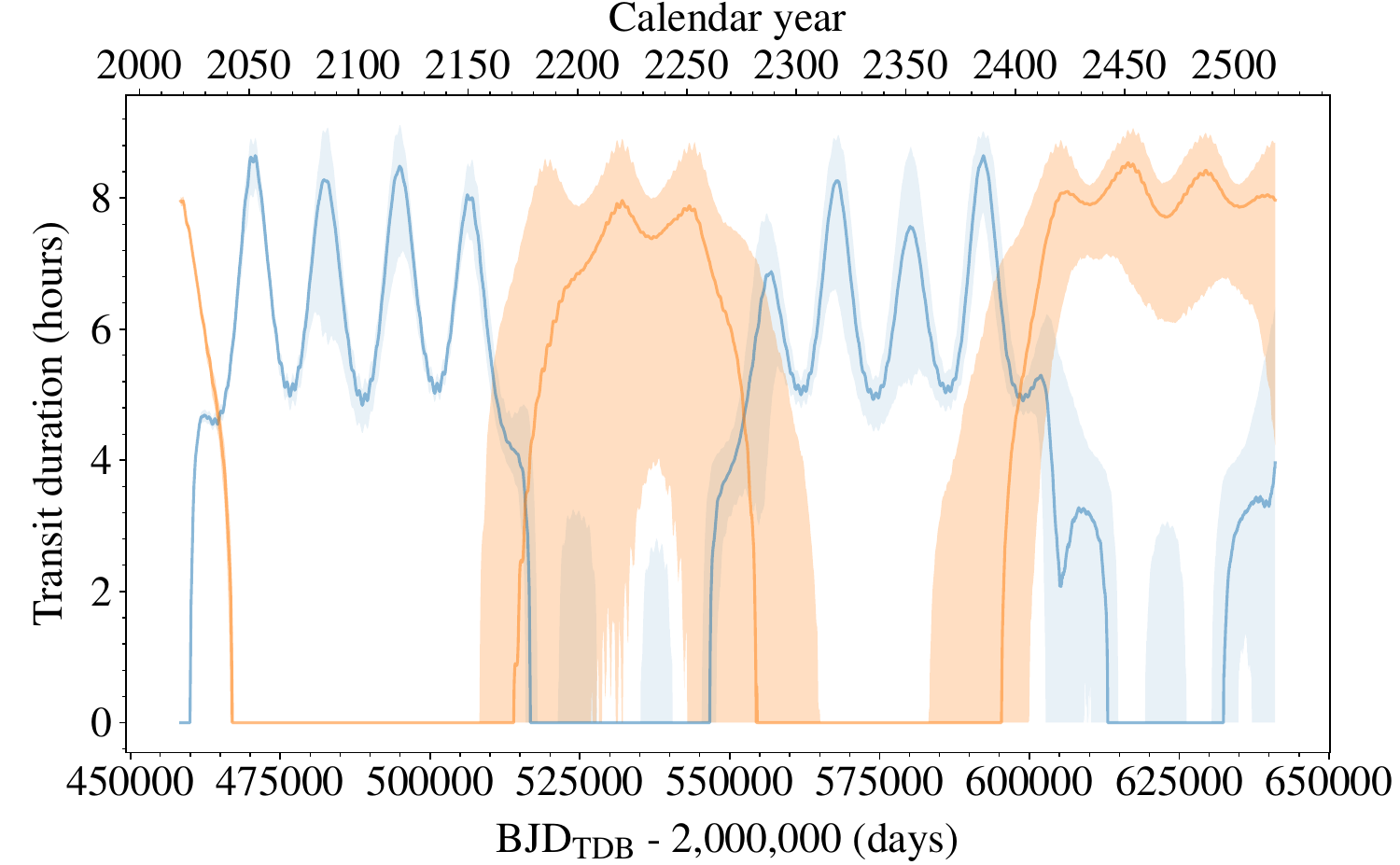}
  \caption{Same as Fig.~\ref{figure:TDVs}, but for the following 500~years. The oscillation on top of the 250~year nodal precession cycle corresponds to the periastron precession at the superperiod (33~years). } \label{figure:TDVsLongTerm}
\end{figure}

\section{Results and discussion}\label{section:results}

The only dynamical constraint we can place on planet~b is an upper mass limit of 100~\MEarth at 95\% confidence. In the following, we therefore concentrate on planets~d and c, the Jupiter-size pair in the 2:1 MMR, for which we have achieved a more detailed characterization, provided that the model is correct. This would no longer hold if the system would contain additional dynamically relevant planets.

The measured masses (2.0 and 2.6~\Mjup) and radii ($\simeq 1$~\Rjup) place both planets in the regime of compact, non-inflated warm Jupiters. Their equilibrium temperatures of 525~K and 413~K are too low for radius-inflation mechanisms to operate efficiently \citep{Demory2011}. Consequently, their densities are consistent with standard cooling--contraction models \citep{Guillot2005,Fortney+2007}. The high surface gravities imply small atmospheric scale heights, and neither planet is an easy target for transmission spectroscopy. Nevertheless, their temperatures fall in an interesting regime for atmospheric chemistry, where cloud formation and the CO--CH$_4$ transition become important \citep{Lodders2002}, possibly allowing the possibility to constrain the planets' internal luminosities \citep[see][]{Welbanks+2024,Sing+2024}. 

The TTVs (Figs.~\ref{figure:TTVs} and \ref{figure:TTVsLongTerm}) show two main periodicities: the resonant period ($P_R\sim2.5$~years) and the superperiod \citep[$P_S\sim33$~years;][]{Lithwick2012}. The origin of these two periods is discussed in Sect.~\ref{section:dynamics}. The peak‑to‑node TTV amplitudes on the timescale of the superperiod (Fig.~\ref{figure:TTVsLongTerm}) reach about 5.2~days for planet~d and about 3.4~days for planet~c (these values respectively correspond to 13\% and 4.1\% of the corresponding orbital periods), whereas the amplitudes over the time span of the observations are about 2~days for both planets. These are among the largest absolute TTV amplitudes known (see Fig.~11 in \citealt{McKee2023} and Fig.~10 in \citealt{Vitkova2025})
The transit forecast from the current modeling shows large uncertainties, reaching up to 28~minutes in 2026. Observing additional transits, which have timing precision of a few minutes, would therefore significantly improve the system parameters.

The orbital planes of planets~d and~c precess with a period of approximately 250~years. Together with their inclinations relative to the line of sight and the measured mutual inclination, this geometry causes planets~d and~c to intermittently cease transiting (see Fig.~\ref{figure:TDVsLongTerm}). Currently, both planets transit. According to the model, planet~d began transiting in late 2022 or early 2023 (although initially with such shallow grazing transit that they were not detectable by TESS) and will keep doing so for the next $\sim$150~years, whereas planet c is expected to stop transiting around 2042 for roughly the following $\sim$130~years. This behavior recurs on the 250-year precession cycle. 

Photodynamical modeling of TOI-4504 provides significantly better constraints on the system parameters than an analysis based only on transit times (TTV analysis): both the planet-to-star mass ratios and the orbital eccentricities are determined with a precision better than 1\%. Eccentricities and planet-to-star mass ratios are determined 52\% to 58\% more precisely with photodynamical modeling than with TTV analysis.
In addition to the transits of planet~d appearing and changing shape, the transit duration variations of planet~c are visible by eye (Fig.~\ref{figure:Transits}); together these constrain the mutual inclination to $2.11\pm0.25$\degree. A TTV analysis of planets~d and~c can, by contrast, only constrain their mutual inclination to $<6.2$\degree\ at 95\% confidence level, and this value corresponds to a lower limit since the inclinations of both planets were fixed at 90\degree\ at $t_{\mathrm{ref}}$. 

\subsection{Comparison with \citet{Vitkova2025}}

\citet{Vitkova2025} correctly identified the 2:1 perturber. However, their reported system parameters differ substantially from ours, with masses of $1.42^{+0.07}_{-0.06}$~\Mjup and $3.77\pm0.18$~\Mjup, and eccentricities\footnote{The eccentricities are osculating values for different reference epochs, but orbital evolution only makes a minor contribution to the discrepancy.} of $0.0445^{+0.0010}_{-0.0009}$ and $0.0320^{+0.0016}_{-0.0014}$ for planets~d and c, respectively. 
Their adopted priors for the parameter space exploration \citep[Table~A1 in][]{Vitkova2025} excluded the solution found in this work, specifically the mass (parameterized by the stellar RV amplitude) and eccentricity (parametrized as $e\cos\omega$, $e\sin\omega$) of planet~d. To explore this further, we broadened their prior on orbital eccentricities and allowed for higher planet‑to‑star mass ratios, finding a multimodal posterior, including one mode compatible with the solution of \citet{Vitkova2025} and another consistent with our own posterior (see Appendix~\ref{section:partial_dataset}). {\it This illustrates the degeneracies inherent in TTV inversions when the perturber is not observed in transit, which can lead to significantly different system parameters all matching the transit times.}

We note that the argument of periastron and the mean anomaly reported in Table~4 of \citet{Vitkova2025} correspond to the stellar orbit \citep{Householder2022}, despite the caption stating that the orbital elements are given in the Jacobi frame. Consequently, to recover their published TTV fit, we added 180\degree\ to those angles before inputting the maximum-likelihood solution into their TTV code \citep[\ttvfast;][]{Deck2014}. 

We also tested the maximum-likelihood solution of \citet{Vitkova2025} using the resonance model outlined in Appendix~\ref{FOT} and found that it corresponds to a system in the non-resonant librating state with $R=1.4$. In particular their average period ratio is 2.061 with associated superperiod 1352~days, while our average period ratio is 2.006, placing the system much deeper in the resonance with superperiod 33.4~years. Note that at around 930~days, the TTV signal varies with the resonant period, not the superperiod as stated by \citet{Vitkova2025}.

\subsection{Relaxed dynamical state and similarities with GJ~876}\label{section:dynamics}

GJ~876 is a non-transiting 4-planet M-dwarf 
system for which the three outer planets (whose masses and periods are $m_c=0.9M_J$, $m_b=2.6M_J$
and $m_e=0.05M_J$, and $P_c=30.1\,{\rm d}$, $P_b=61.1\,{\rm d}$ and $P_e=123.8\,{\rm d}$) 
participate in a 4:2:1 Laplace resonance 
\citep{Delfosse1998,Marcy1998,Marcy2001,Rivera2005,Rivera2010,Correia2010,Millholland2018,Moutou2023}.
\citet{Lee2002} recognized that the unusual alignment of the orbits of the giant pair c and b and their significant eccentricities
could not be understood with simple linear theory.
Here we use a new formulation of the problem (Mardling {\it in prep}; hereafter M1 and summarized in 
Appendix~\ref{FOT}) to highlight and understand
some of the striking similarities between the giant pairs orbiting 
TOI-4505 and GJ~876, and discuss the implications for the formation of resonant giants.

To avoid confusion regarding the order of the orbital periods of the pair TOI-4505d and TOI-4505c, and similarly of the pair
GJ~876c and GJ~876b, in this section we will use the subscripts ``i'' and ``o'' to indicate ``inner'' and ``outer''
for the two giant planets in 2:1 resonance. 

The dynamical state of a multiplanet system is of particular interest for the constraints it places on the system's formation
and subsequent evolution. In particular, the period ratios, eccentricities and orbital orientations provide evidence
for planet-disk interaction when they are close to one of the relaxed states predicted by migration theory and simulations;
if they are not, further insight is required. 
Planet-planet scattering is often invoked, either at the time of formation \citep[eg.][]{Rasio1996} or as a consequence of 
long-term instability \citep[eg.][]{Kokubo2025,Izidoro2017}. Evidence for the latter is arriving with the characterization
of very young systems \citep{Dai2023,Dai2024}.

With an increasing number of transiting systems being accurately characterized using photodynamical analysis,
it is becoming clear that the architecture of low-mass planetary systems often seem to be
inconsistent with simple migration theory.
Especially curious are systems for which the forced component of the eccentricities are 
damped to their non-zero stationary values
as expected, while the free components are not; these states are called ``supercycles'' in M1 
(see Appendix~\ref{FOT} for new
definitions of forced and free eccentricity which are valid for all configurations close to a first-order commensurability,
and the various supercycle systems listed there under point (x)).
In this respect it will be particularly interesting to see if the free components of young systems follow suit once
they are accurately characterized. In contrast, the giant pairs in TOI-4504 and GJ~876 {\it do} seem to 
conform with formation theory with little subsequent evolution,
a remarkable fact given their estimated ages of $10.0^{+2.9}_{-3.6}\,{\rm Gyr}$ and 0.1-5 Gyr respectively \citep{Vitkova2025,Correia2010}.

We begin with an overview of the relaxation process according to simple migration theory \citep{Papaloizou2000,Goldreich2014},
using the first-order (in eccentricity) formulation of M1 to describe the possible final states
(Appendix~\ref{FOT}). Note that while first-order Hamiltonian formulations 
are available \citep[e.g.,][]{Batygin2015,Nesvorny2016,Petit2017}, M1 uses a general dynamical systems approach
which provides physical intuition and remains close to the familiar orbital elements.

While the process of migration tends to damp a single planet's orbital eccentricity completely \citep[at least when the disk
is not eccentric,][]{Kley2012}, for two planets converging towards
a low-order commensurability, an eventual balance between the difference in disk torques and the 
planet-planet torque leaves the pair with a constant period ratio and constant 
non-zero eccentricities, or finds its equilibrium in a limit-cycle state 
in which these quantities librate around a fixed-point \citep[``overstable libration'',][]{Goldreich2014}.
Moreover, after such a state is reached, it persists as the planets continue to migrate until the disk disappears. 
This assumes the latter occurs before the system reaches the disk edge as seems to be the case for 
the giant pairs in TOI-4504 and GJ~876; here material between the planets is likely to have been
completely absent by the time resonance capture occurred due to gap clearing, with convergent migration being 
outward for the inner planet and inward for the outer planet \citep{Bryden2000,Kley2000}.

If the additional acceleration of the planets produced by the disk can be characterised by a
tangential and radial component such that its form is
\be
{\bf a}=-\tau_{a}^{-1}\dot{\bf r}-2\tau_{e}^{-1}\left(\dot{\bf r}\cdot\hat{\bf r}\right)\hat{\bf r},
\label{diss}
\ee
where ${\bf r}$ is the position of the planet relative to the star (or to the centre of mass of the star-planet system
interior to the orbit) and $\tau_a$ and $\tau_e$ are the migration and eccentricity damping timescales
\citep{Papaloizou2000}, then the equilibrium eccentricities will be $\sim\sqrt{\tau_e/\tau_a}$
\citep{Goldreich2014} and the distance to exact commensurability $\sim (m_{\rm co}/m_*)\sqrt{\tau_a/\tau_e}$ (M1),
where $m_{\rm co}$ is the mass of the companion planet (see Eq.~\rn{eplus}).
Therefore the closer the 
two damping timescales are, the higher the eccentricities and the smaller the distance to exact commensurability.
\citet{Papaloizou2010} adopt a value of $\tau_a/\tau_e\simeq 400$ for Type~I migration which leads
to eccentricities of $\sim 0.05$ \citep[more like 0.02 when appropriate factors are included; see Eq.~(24) of][]{Goldreich2014}.
While such eccentricities are typical in chains of low-mass planets, 
they are considerably smaller than those of the inner giants of TOI-4504 and GJ~876 which are 0.29 and 0.26 respectively.
Moreover, both
are close to or at their equilibrium values (see below), suggesting that $\tau_a$ and $\tau_e$ have similar values for
Type~II migration of pairs of giants (at least for these two systems; see \citealt{Kley2005} for a discussion
of migration scenarios for GJ~876).

In addition to non-zero relaxed eccentricities,
the resonance-capture process results in orbital anti-alignment when the equilibrium eccentricities are small,
as is easily demonstrated using first-order theory (see Appendix~\ref{FOT} and Eq.~\rn{eplus}). 
The theory also shows that the 
associated eccentricity vectors (expressed as complex numbers and involving the mean longitudes)
can be decomposed into ``forced'' (or {\it resonant}) and ``free'' (or {\it secular})
components (Appendix~\ref{FOT}, point (viii)),\footnote{Note
that the definitions of forced and free eccentricity given in M1 and Appendix~\ref{FOT} are different to those in
\citet{Lithwick2012}, the latter formulated for non-librating systems (often referred to as being ``outside the resonance'',
although the phase space need not be resonant for a system to be librating). 
In particular, for M1 the forced component is associated with energy and angular momentum exchange 
between the orbits which involves variations of both the eccentricities and semimajor axes with associated 
resonant period $P_R$, while the free part
is constant at first-order in eccentricity \citep[see Sect.~\ref{LXW} for the meaning of ``free'' in][]{Lithwick2012}.
A consequence of this is that TTVs only become sensitive to the free
components at second order so that for systems with small eccentricities there is considerable degeneracy
in their photodynamical modelling \citep[a good example of this phenomenon is \mbox{K2-19},][]{Almenara2025}.
That the eccentricities of the giant pair in \mbox{TOI-4504} can be measured from the TTVs with such small uncertainties
is a direct consequence of the fact that the free eccentricities vary due to nonlinear interaction between the
forced and free modes.}
and that while the forced eccentricities
relax to their equilibrium values as discussed above, the free components relax to zero (M1).

\begin{figure*}[t]
  \includegraphics[width=0.95\textwidth]{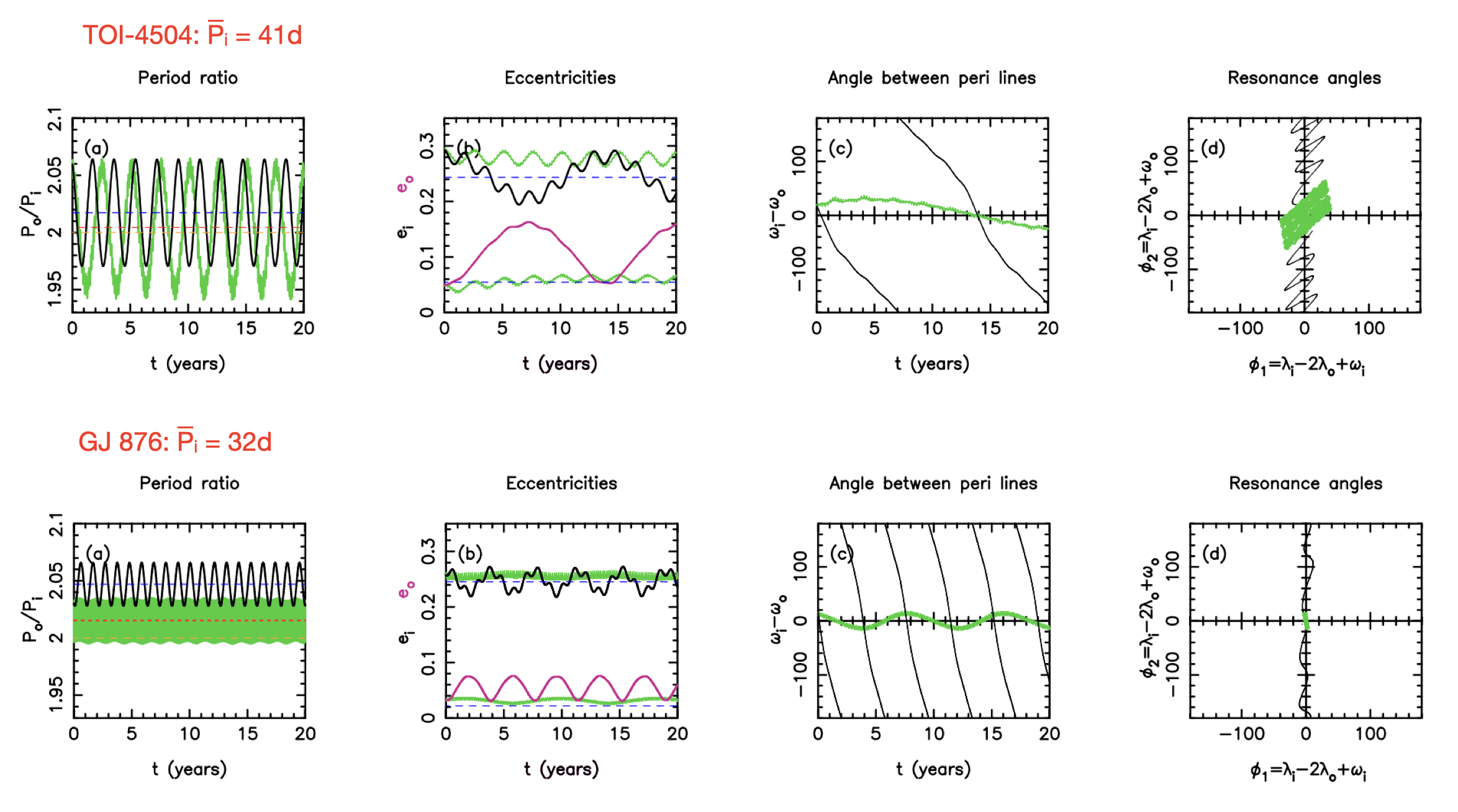}
  \caption{Dynamical evolution of TOI-4504 and GJ~876.
Green curves are $N$-body integrations with initial conditions from the  photodynamical MAP solution for TOI-4504,
and the elements at epoch for the best-fit (radial velocity) coplanar four-planet fit of \citet{Millholland2018} for GJ~876.
The black and pink curves show first-order theory solutions to the system (Eq.~\rn{ODEs}),
with $P_{\rm o}/P_{\rm i}=(n+1)/n+(1+\theta')\Delta\sigma$,
$\ei=[z_1z_1^*]^{1/2}$ with $z_1$ from Eq.~\rn{z1ff} and similarly for $\eo$ via $z_2$,
the resonance angles from the phases of $z_1$ and $z_2$ (see Eq.~\rn{resecc}), 
and $\varpi_{\rm i}-\varpi_{\rm o}=\phi_1-\phi_2$. 
The blue-dashed lines correspond to first-order theory equilibria, the orange-dashed line in (a) to exact commensurability,
and the red-dashed line in (a) to the average value of the true period ratio.
Notice that the true $\phi_1$, $\phi_2$ and $\varpi_{\rm i}-\varpi_{\rm o}$
all librate around zero, consistent with linear theory for $\phi_1$ because
$\ei^{(+)}>\ei^{({\rm free})}$, but not for $\phi_2$ for which theory predicts
circulation because $\eo^{(+)}<\eo^{({\rm free})}$,
and therefore circulation of $\varpi_{\rm i}-\varpi_{\rm o}$ (Appendix~\ref{FOT}).
See text for discussion regarding the current evolution of the eccentricities which reveals that both systems are close 
to the fully relaxed state.
} \label{figure:compare}
\end{figure*}

\begin{figure}
  \centering
  \includegraphics[width=0.49\textwidth]{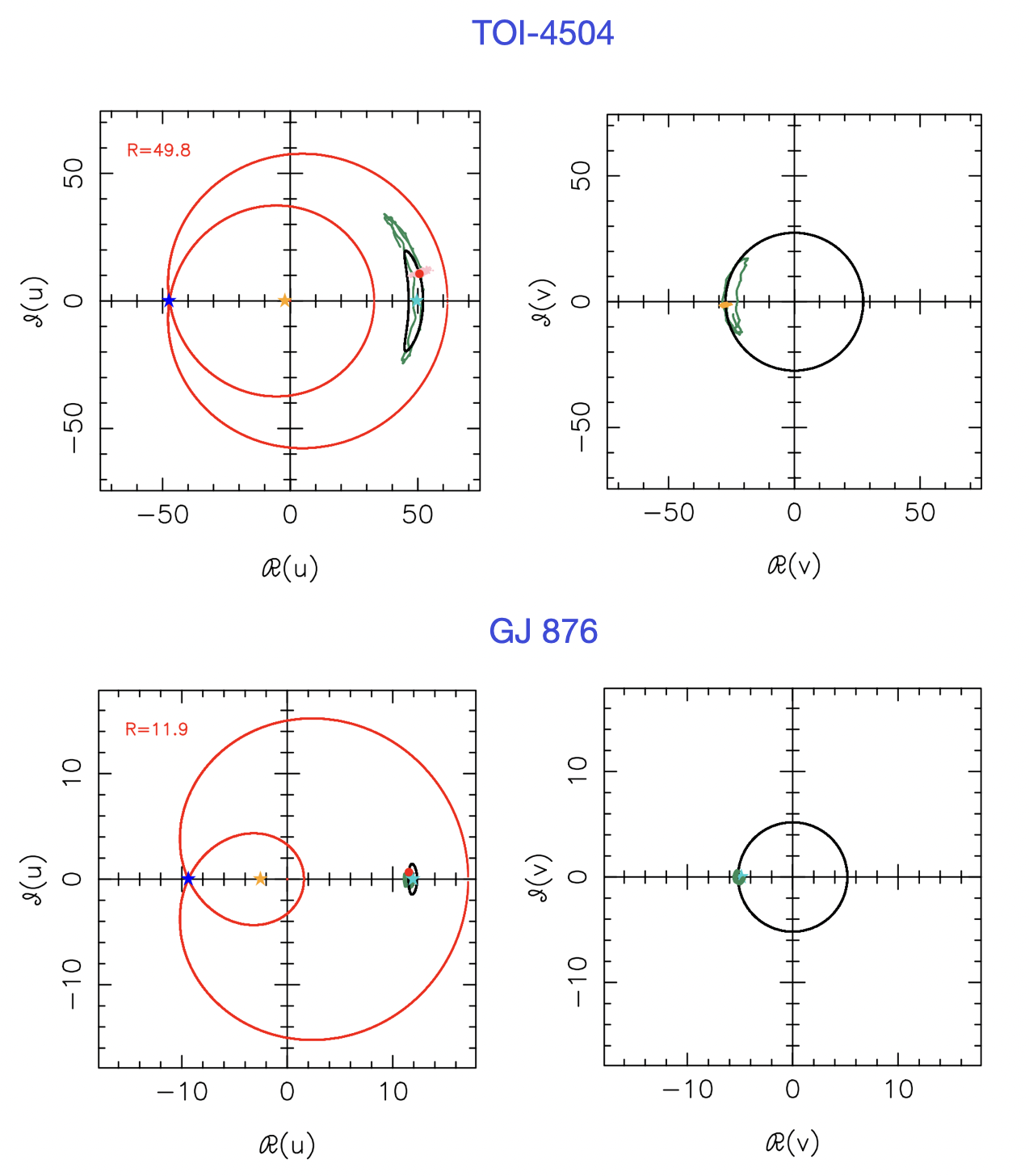}
  \caption{Phase plots for TOI-4504 and GJ~876, with initial conditions for GJ~876 from \citet{Millholland2018}.
  The axes are the real and imaginary parts of $u$ and $\upsilon$, defined in Eqs.~\rn{u} and \rn{v}.
  Green curves are $N$-body solutions while black curves are from first-order theory.
Note that for both systems $|\upsilon|$ is a significant fraction of $R$, indicating the strength of 
interaction between the resonant and secular modes.
Both systems are close to their primary fixed points and are likely to be fully relaxed in limit-cycle states;
see text for detailed discussion.
 Notice different scales for TOI-4504 and GJ~876.} 
  \label{figure:phase}
\end{figure}

Noting that our photodynamical analysis yields small uncertainties for the eccentricities (Table~\ref{table:results}),
Table~\ref{table:dynamical} 
\begin{table*}
\tiny
  \setlength{\tabcolsep}{4pt} 
\renewcommand{\arraystretch}{0.95}
\caption{Dynamical parameters of TOI-4504 and GJ~876.}             
\label{table:dynamical}      
\centering                          
\begin{tabular}{l c c c c c c c c}        
\hline\\\smallskip
System & $\overline P_{\rm o}/\overline P_{\rm i}$ & $e_{\rm i}$,\, $e_{\rm i}^{(+)}$, $e_{\rm i}^{({\rm free})}$ 
& $e_{\rm o}$, \, $e_{\rm o}^{(+)}$, $e_{\rm o}^{({\rm free})}$ & $\langle\omega_{\rm i}-\omega_{\rm o}\rangle$,\,
$(\langle\phi_1\rangle,\langle\phi_2\rangle)$ & $m_*/M_\odot$ & $m_i/m_*$, $m_o/m_*$, $m_o/m_i$ & $R$ & $P_S$, $P_R$ (yr)
 \\    
\hline                        
&\\
TOI-4504 &   2.006 [2.017] &   0.293, [0.243], [0.038] & 0.050, [0.054], [0.106] &   0, \hspace{0.5cm} (0,0) 
& $0.88$ & 0.002, 0.003, 1.3  & 172 [50] & 33.4, 2.5 \\  
&    &&&&&&&     [13.6],  [1.8] \\
&\\
GJ~876 &  2.0174 [2.047]  & 0.257, [0.246], [0.019] & 0.033, [0.022], [0.053] &  0, \hspace{0.5cm} (0,0) &  $0.33$ & 
0.002, 0.008, 3.2  & 21 [12]  & 7.3, 1.5 \\   
&    &&&&&&&     [3.8],  [1.0] \\
\hline                                   
\end{tabular}
\tablefoot{First-order theory values are shown in square brackets, with
$e_{\rm i}^{(+)}/e_{\rm o}^{(+)}=(\mo/\mi)\alpha^{-1/2}|\mv_1|/\mv_2=3.50(\mo/\mi)$ and 
$e_{\rm i}^{({\rm free})}/e_{\rm o}^{({\rm free})}=\mv_2/|\mv_1|=0.36$ for 2:1.
Notice that $e_{\rm i}\simeq e_{\rm i}^{({\rm free})}+e_{\rm i}^{(+)}$ and
$e_{\rm o}\simeq e_{\rm o}^{({\rm free})}-e_{\rm o}^{(+)}$ (see (b) panels of Fig.~\ref{figure:compare}). }
\end{table*}
lists various relevant dynamical parameters for the giant pairs in 
TOI-4504 and GJ~876, while Figs.~\ref{figure:compare} 
and \ref{figure:phase} 
compare their evolution over 20~years.
The dynamical states of the two systems are remarkably similar, with each having:
\begin{enumerate}
\item[(i)]
Inner eccentricities $\sim 0.25-0.3$;
\item[(ii)]
Similar inner planet-to-star mass ratios (in spite of significantly different stellar masses);
\item[(iii)]
Both resonance angles 
(and hence the difference in the longitudes of periastron) librating around zero;
\item[(iv)]
A phase space position deep in the 2:1 resonance and close to the fully-relaxed state (Fig.~\ref{figure:phase});
\item[(v)]
Significantly non-zero {\it relaxed} free eccentricity (Fig.~\ref{figure:phase}); this aspect is especially
interesting for the discussion below and is the reason it is possible to measure the eccentricities and
apsidal angles with such low uncertainties.
\end{enumerate}
Figure~\ref{figure:compare} compares the predictions of first-order theory (black and pink curves) with
$N$-body integrations (green curves). Two periods are clearly evident.
The shorter resonant period, $P_R$, governs the variation of the
period ratio which theory underestimates
by a factor of 1.4-1.5, while the difference in the apsidal angles is dominated by the superperiod, $P_S$, the true value
of which is underestimated by a factor of 2-2.5. Both periods are obvious in the theoretical variation of 
the eccentricities, with large amplitude variations associated with the superperiod, 
while the true variation is almost exclusively governed by the resonant frequency for
TOI-4504 (with amplitude well-predicted by theory). The resonant frequency is almost absent from the true
eccentricities of GJ~876. For
both systems the true eccentricities are such that their average values are
\be
\langle e_{\rm i}^{N}\rangle\simeq e_{\rm i}^{({\rm free})}+e_{\rm i}^{(+)}
\hand
\langle e_{\rm o}^{N}\rangle\simeq e_{\rm o}^{({\rm free})}-e_{\rm o}^{(+)},
\label{aveccs}
\ee
where the superscript $N$ denotes $N$-body,
$e_{\rm i}^{(+)}$ and $e_{\rm o}^{(+)}$ are the inner and outer primary equilibrium eccentricities
predicted by linear theory (blue dashed lines), 
and $e_{\rm i}^{({\rm free})}$ and $e_{\rm o}^{({\rm free})}$ are the corresponding free eccentricities. 
That $\langle e_{\rm i}^{N}\rangle$ and $\langle e_{\rm o}^{N}\rangle$ can be expressed in this way is explained below.

While the free eccentricities are constant at linear order (as are their associated apsidal angles),
and moreover are zero in the fully relaxed state, at higher order this is no longer true. In fact given the  
high eccentricities of TOI-4504 and GJ~876, themselves a result of the close proximity of the average period ratio
to exact commensurability,
nonlinear interaction between the resonant and secular modes is 
significant enough to flip the equilibrium value of the resonance angle
$\phi_2$ from $\pi$ to zero (with $\phi_1$ remaining at zero),
so that the difference in apsidal angles
librates around zero rather than $\pi$ (panels (c) and (d) of Fig.~\ref{figure:compare}), and to make the relaxed value
of the free eccentricity non-zero and the associated free apsidal angles (which are the same
for both planets) a function of time.  
{\it This latter fact makes the free eccentricity visible to the TTVs, which in turn allows for precise 
determination of the full eccentricity vectors as long as the observing baseline covers enough of the superperiod.} 

Moreover, since the values of the phase-space quantities $u$ and $\upsilon$ (defined in Eqs.~\rn{u} and \rn{v} and plotted in Fig.~\ref{figure:phase}) are non-zero and almost constant for TOI-4504 and GJ~876,
we have from Eqs.~\rn{uandv}, \rn{psi} and \rn{wpsi} that 
\be
\langle\dot\varpi^{({\rm free})}\rangle
=\langle\dot\varpi^{({\rm forced})}\rangle
=-\langle\dot\lambda_n\rangle
\equiv\nu_S=2\pi/P_S,
\ee
where $-\lambda_n$ is the longitude of conjunctions defined in Eq.~\rn{lambdan}.
Thus the average rates of advance of the forced and free apsidal lines follow the line of conjunctions
and are equal to the superfrequency,
which by Eqs.~\rn{resecc}, \rn{z1ff} and \rn{z2ff} implies that the apsidal lines themselves follow the line of conjunctions,
that is,
\be
\langle\dot\varpi_{\rm i}\rangle
=\langle\dot\varpi_{\rm o}\rangle
=\nu_S.
\ee

Note that if TOI-4504 had been precisely at the primary fixed point (the aqua stars in Fig.~\ref{figure:phase}), only 
the superperiod $P_S\simeq 33.4$~yr would be present in the TTV signal;
that it is not means that the much shorter resonant period $P_R=2.5$~yr is also present,
allowing for full characterization of the system on a 7~year baseline.

The expressions \rn{aveccs} are consistent with the M1 definition of a supercycle state for which
the forced components of the eccentricity variables $z_1$ and $z_2$ reside at a fixed point 
and the free components are non-zero. Such a state implies
that at the time of resonance capture the damping timescale of the forced component 
was shorter than that of the free component
(see point (x) in Appendix~\ref{FOT} which includes a list of
observed systems close to the supercycle state).
For our purposes here, the linear theory predicts that when 
$e_{\rm i}^{({\rm free})}/e_{\rm i}^{(+)}<1$ and
$e_{\rm o}^{({\rm free})}/e_{\rm o}^{(+)}>1$, as is the case for both TOI-4504 and GJ~876, the maxima
and minima of the variation of the eccentricities are such that
\be
e_{\rm i}^{({\rm max,min})}=e_{\rm i}^{(+)}\pm e_{\rm i}^{({\rm free})}
\hand
e_{\rm o}^{({\rm max,min})}= e_{\rm o}^{({\rm free})}\pm e_{\rm o}^{(+)}.
\ee

But in the linear theory, non-zero free eccentricity implies that the system is not fully relaxed,
while at higher orders the relaxed values of the free eccentricities are not zero.
Although such relaxed values tend to be much smaller than any residual true free eccentricity, for TOI-4504 and GJ~876
they are significant, especially for the outer planets (Table~\ref{table:dynamical}). 
The black (first-order theory) and green ($N$-body) curves in the 
top right-hand panel of Fig.~\ref{figure:phase} shows the evolution
of the real and imaginary components of the secular phase-space
quantity $\upsilon$ (Eq.~\rn{v}) for TOI-4504, whose magnitude is proportional
to the free eccentricities (Eq.~\rn{ebfree}). Also shown in orange are 7000 posterior samples for this quantity,
revealing that $\upsilon$ is tightly constrained.
In the linear theory the magnitude of $\upsilon$ (the radius of the black circle) is constant, while its completely relaxed
value is zero. In contrast the green curve shows the actual evolution of this quantity, with libration around a negative
real value, the latter being the true equilibrium value of $\upsilon$ (and hence giving the true free eccentricities
via Eq.~\rn{ebfree}). 
{\it The fact that the $N$-body MAP solution and posterior values are near this value tells us that 
the TOI-4504 system is close to or actually fully relaxed} (if in a limit-cycle state;
see point (ix)(B) in Appendix~\ref{FOT}), {\it implying very little evolution since formation.}
Moreover, the fact that $\upsilon$ remains close to its negative equilibrium value over 
a libration cycle\footnote{Since there is nonlinear interference between the $u$ and $\upsilon$ modes, 
both the resonant and super frequencies influence their variation; 
the libration cycle shown here has period $P_R$.}
explains why the true amplitudes of
the super components of the eccentricities are small (green curves in panel (b) of Fig.~\ref{figure:compare}), and
why the average values of the true eccentricities are given by Eq.~\rn{aveccs}.

The corresponding black and green curves in the top left-hand panel of Fig.~\ref{figure:phase} shows the evolution
of the real and imaginary components of the resonant phase-space quantity $u$ (Eq.~\rn{u})
for TOI-4504, together with the separatrix
(red curve), the primary fixed point (aqua star), the hyperbolic fixed point (blue star), and
7000 posterior samples for $u$ (pink dots), again revealing that this quantity is tightly constrained.
The red dot is the MAP value of $u$.
Just as for $\upsilon$, $u$ remains close to its equilibrium value, however its variation is responsible for 
TTV detection over the 7 year observing
baseline; again had the system been in the stationary state the $P_R=2.5$ yr resonant period
would not be present in the signal, leaving only the $P_S=33$ yr period.

The bottom two panels of Fig.~\ref{figure:phase} show the same analysis for GJ~876, revealing that it is even
closer to its stationary state than TOI-4504.

Finally, as outlined in the Appendix the linear theory can be used to estimate the resonant and superperiods;
these are listed in square brackets in Table~\ref{table:dynamical} for both TOI-4504 and GJ~876. But the true
periods are considerably longer, a result of the fact that the average value of the period ratio is closer
to exact commensurability than predicted by linear theory. As a consequence, both systems are
deeper in the 2:1 resonance than suggested by the already high linear-theory values
of the resonance parameter $R$ of 50 and 12 respectively.
Estimates of the true values of $R$ are given in Table~\ref{table:dynamical}, 
obtained using the true values of $P_R$ and $P_S$
in Eq.~\rn{PSR} with the ansatz that this relationship is
valid generally for systems close to the primary fixed point.

\subsection{Summary of impact of nonlinear effects on the TTVs}

When the eccentricities of a two-planet system are small and the free eccentricities are non-zero, both the free eccentricities
and apsidal angles are constant at first order and slowly varying at higher order (hence the designation ``secular'';
see Appendix~\ref{second} for a discussion of second-order effects).
The TTVs are therefore largely insensitive to this component of the eccentricity vectors in this case, producing
extensive degeneracy in the eccentricity and apsidal angle projections 
of the posterior (see Fig.~6 in \citealt{Almenara2025} for an example of this phenomenon).

In contrast, when the eccentricities are significant because of a system's extreme proximity to exact commensurability,
a situation likely to be a result of the similarity of the migration and eccentricity damping timescales during Type-II migration, 
nonlinear interaction between the resonant and secular modes causes the free apsidal lines to evolve,
making the free components visible to the TTVs. As a result the eccentricity-orientation degeneracy is lifted,
allowing for accurate determination of these quantities (as long as the observing baseline covers enough of the variation,
and the photometric precision is adequate).

We again point the reader to Appendix~\ref{FOT} which presents a self-contained summary of 
the new and accessible formalism of M1 for the study of systems near a first-order commensurability,
whether they be librating or circulating, resonant or non-resonant. 
The analysis in turn yields simple
expressions for the dominant resonant and super harmonics of the TTVs in all these cases (point (vii) in the summary). 
Note that
if both these harmonics (whose associated frequencies are in general non-commensurate) are present in the signal,
there is enough information to determine the planet-to-star mass ratios, independent of any degeneracy in 
the eccentricities and apsidal angles (again see  \citealt{Almenara2025} where this is demonstrated).

In addition, new general definitions of forced and free eccentricity are introduced which are valid for 
both librating and circulating systems, offering particular insight into the relaxed states of TOI-4504 and GJ~876.
These are compared to the definitions of \citet{Lithwick2012}
which only hold for circulating systems sufficiently far from the libration-circulation boundary.

Finally, Appendix~\ref{FOT} discusses various possible relaxed states, illustrating these with well-characterized 
systems including 
TOI-1408,
TOI-216,
TOI-7510,
K2-19,
K2-266,
Kepler-18,
Kepler-36,
Kepler-88,
Kepler-9 and the
Neptune-Pluto system.
Note that the formalism provides a simple way to predict the librating
or circulating behaviour of the resonance angles when a system is in the librating state (be it resonant or not).

\bigskip

As noted by \citet{Vitkova2025}, TOI-4504 will be observed by PLATO \citep{Rauer2014}, whose observations are scheduled to begin in 2027. This will establish it as one of the most precisely characterized planetary systems hosting giants, especially from the point of view of their resonant dynamical evolution.


\begin{acknowledgements}

This paper includes data collected by the TESS mission. Funding for the TESS mission is provided by the NASA's Science Mission Directorate.

Resources supporting this work were provided by the NASA High-End Computing (HEC) Program through the NASA Advanced Supercomputing (NAS) Division at Ames Research Center for the production of the SPOC data products.

This paper includes data collected by the TESS mission that are publicly available from the Mikulski Archive for Space Telescopes (MAST).

This work is based on observations collected with EulerCam mounted on the 1.2~m Swiss Euler telescope at La Silla Observatory, Chile.

Simulations in this paper made use of the \rebound code which can be downloaded freely at \url{http://github.com/hannorein/rebound}. 

These simulations have been run on the {\it Bonsai} cluster kindly provided by the Observatoire de Gen\`eve.

A.L., J.K, and J.M.A. acknowledges support of the Swiss National Science Foundation under grant number TMSGI2\_211697.
M.L. acknowledges support of the Swiss National Science Foundation under grant number PCEFP2\_194576.
TF acknowledges funding from the French ANR under contract number ANR24-CE493397 (ORVET), and the French National Research Agency in the framework of the Investissements d’Avenir program (ANR-15-IDEX-02), through the
funding of the “Origin of Life” project of the Grenoble–Alpes University. 

This work has been carried out within the framework of the NCCR PlanetS supported by the Swiss National Science Foundation under grant 51NF40\_205606.

\end{acknowledgements}

\bibliographystyle{aa}
\bibliography{TOI-4504}

\begin{appendix}

\FloatBarrier

\section{Parameter space exploration with the partial dataset of \citet{Vitkova2025}}\label{section:partial_dataset}

\begin{figure}
  \centering
  \includegraphics[width=0.49\textwidth]{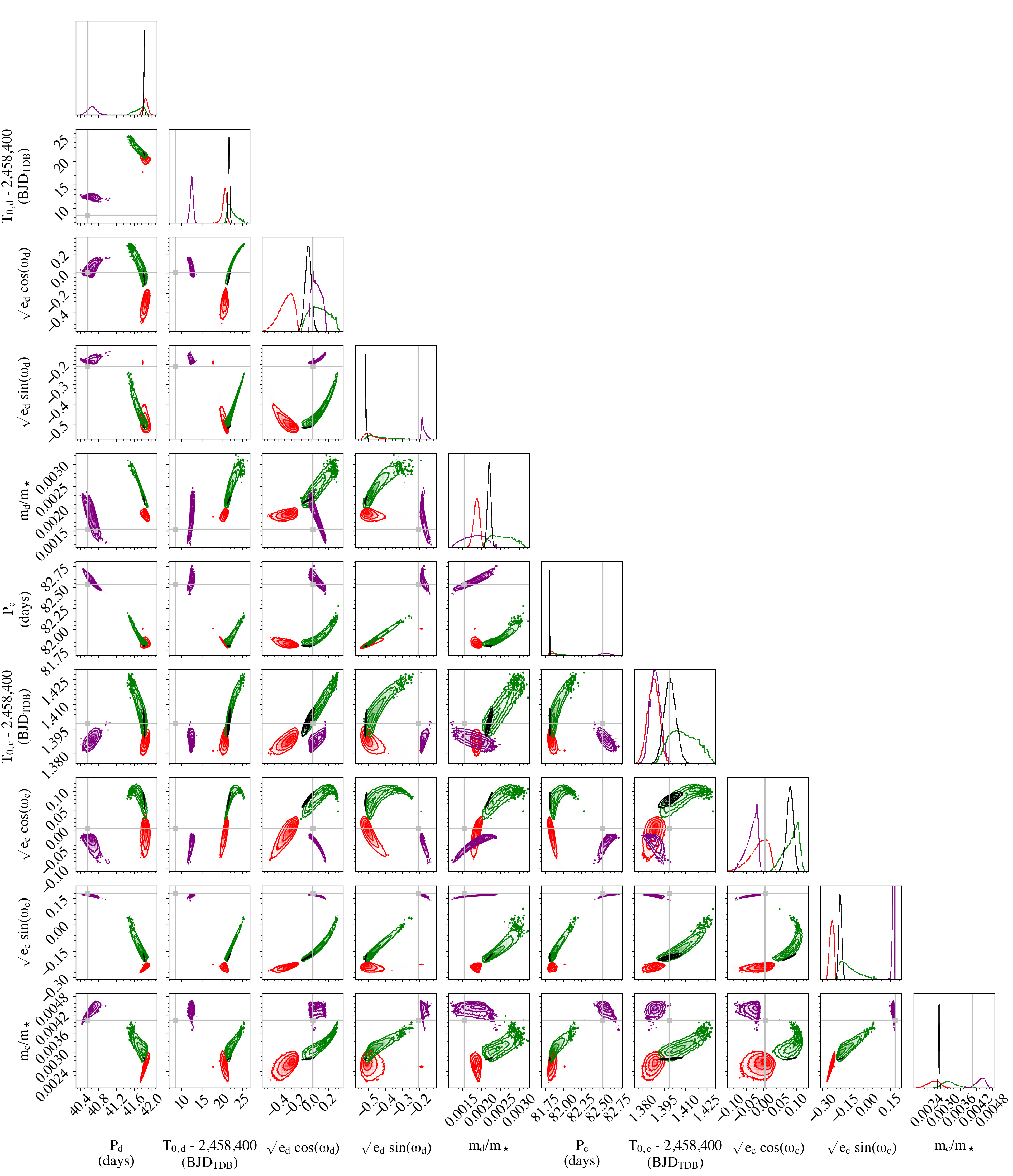}
  \caption{One- and two-dimensional projections of the posterior sample \citep{corner} for the jump parameters of the partial-dataset modeling, showing the different modes identified (purple, green, and red). The black posterior corresponds to the full dataset. The solution of \citet{Vitkova2025} is indicated in gray. These parameters have a different definition and reference epoch from those used in Sect.~\ref{section:analysis} and cannot be compared directly.} \label{figure:nautilus_corner}
\end{figure}

\begin{figure}
  \centering
  \includegraphics[width=0.49\textwidth]{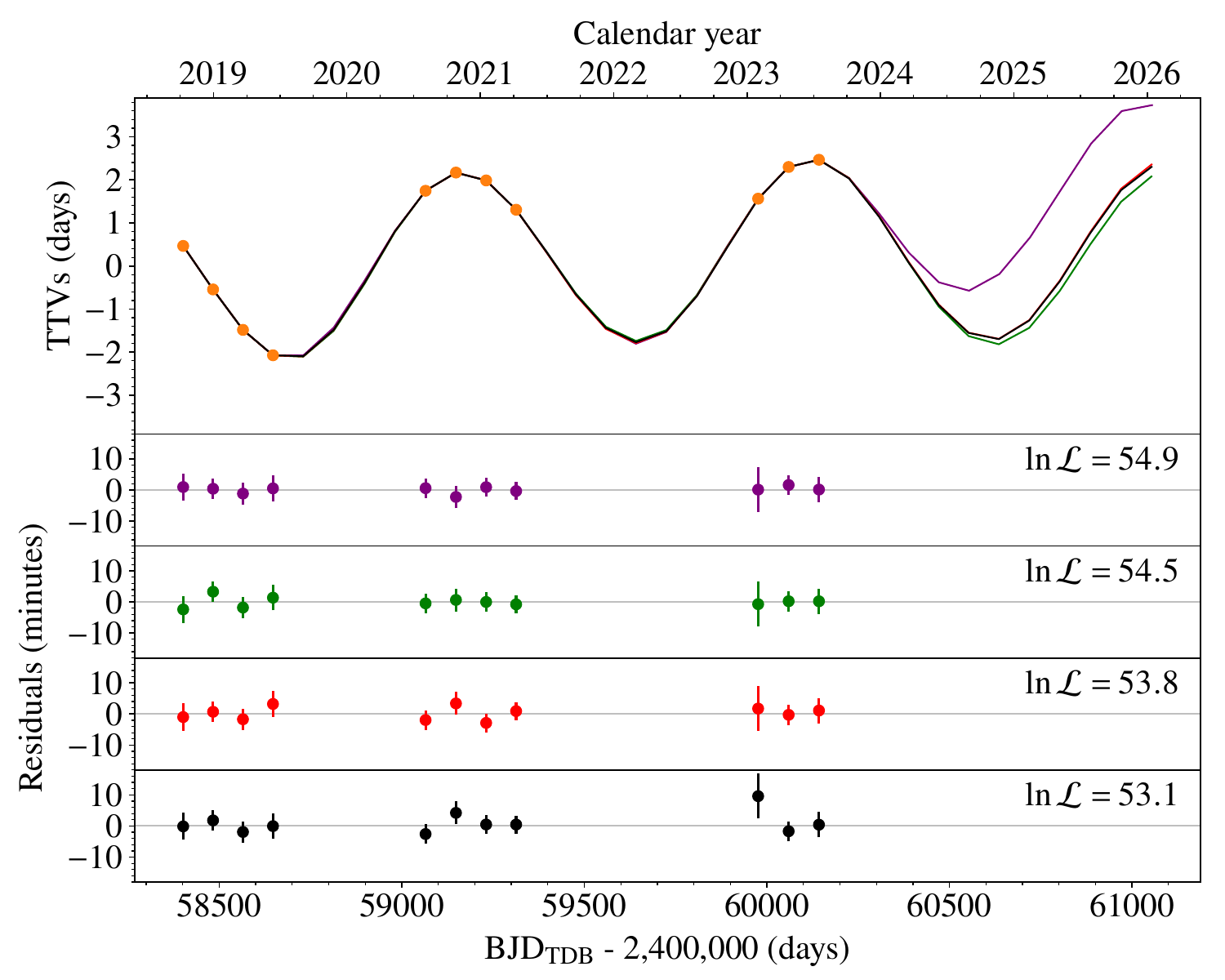}
  \caption{TTV model and residuals for the maximum-likelihood solution of each mode shown in Fig.~\ref{figure:MAP_partial_dataset} (same color code). The corresponding log-likelihood ($\ln \mathcal{L}$) value is indicated in the residual panel of each model.} \label{figure:MAP_partial_dataset}
\end{figure}

We explored the parameter space around the 2:1 resonance using only the transit times of planet~c (Table~\ref{table:transit_times}) over the same time span available to \citet{Vitkova2025}, that is, epochs 0 to 21 (partial dataset). We modeled the transit times with the \ttvfast code \citet{Deck2014}, adopting a time sampling of 0.1~days and the same reference epoch ($2\,458\,400$~BJD$_{\mathrm{TDB}}$) as \citet{Vitkova2025}. We considered only planets d and c, assuming coplanar orbits, and explored the parameter space with the nested‑sampling \nautilus code \citep{Lange2023}. We found a multimodal posterior (see Fig.~\ref{figure:nautilus_corner}). All these modes provide and adequate fit to the transit times (see Fig.~\ref{figure:MAP_partial_dataset}). The mode shown in purple, compatible\footnote{The match is not exact because of the coplanarity assumption. This difference reflects the additional freedom and the associated complexity introduced when non‑coplanar orbits are allowed.} with the solution of \citet{Vitkova2025}, lies outside the resonance, whereas the red and green modes lie inside the resonance. To obtain a ground truth, we repeated this analysis using all available transit times (Table~\ref{table:transit_times}), including those of planet~d (full dataset), and found a single mode (black in the plots). The green mode encompasses the posterior of the full dataset.

\onecolumn

\section{First-order theory}\label{FOT}

The following provides a summary of the aspects of the first-order theory used in this study
(with ``first-order'' referring to the eccentricities), the full
formulation being presented in M1. Using a dynamical systems approach with the fixed points of 
the phase space serving as reference points, the aim is to keep
as close as possible to familar orbital elements and TTV observables.
The formalism is valid for both librating and circulating systems which are close to the period ratio
$P_{\rm o}/P_{\rm i}\simeq (n+1)/n$, where $n$ is an integer,
either of which may reside in a resonant or non-resonant phase space.
The concepts introduced are illustrated using various well-characterized systems (highlighted in bold below).

In terms of the usual elements, the eight dimensionless
quantities governing a coplanar system close to a first-order commensurability are
$P_{\rm o}/P_{\rm i}$, $n\lambda_{\rm i}-(n+1)\lambda_{\rm o}$, $\ei$, $\eo$, $\varpi_i$ and $\varpi_o$,
as well as the planet-to-star mass ratios $\mi/m_*\equiv\overline m_{\rm i}$ and $\mo/m_*\equiv\overline m_{\rm o}$.
The elements are first replaced by the two complex quantities
\be
z_1=e_b\,{\rm e}^{i\phi_1}
\hand
z_2=e_c\,{\rm e}^{i\phi_2},
\label{resecc}
\ee
where
\be
\phi_1=n\lambda_{\rm i}-(n+1)\lambda_{\rm o}+\varpi_{\rm i}
\hand
\phi_2=n\lambda_{\rm i}-(n+1)\lambda_{\rm o}+\varpi_{\rm o}
\label{phi}
\ee
are the resonance angles,
and the two real quantities $\tilde\theta$ and $\dot{\tilde\theta}$, defined such that minus the longitude of conjunctions is
\be
\lambda_n(t)\equiv n\lambda_{\rm i}-(n+1)\lambda_{\rm o}
=\lambda_n(\tref)+\nu_S(t-t_{\rm ref})
+\tilde\theta(t)-\tilde\theta(\tref).
\label{lambdan}
\ee
Here $\nu_S=2\pi/P_S$ 
is the superfrequency, $t_{\rm ref}$ is the reference time, and $\tilde\theta(t)$ captures the exchange of
energy and angular momentum between the orbits via the
resonant oscillation (with period $P_R$) around the mean increase (or decrease) of $\lambda_n$. 
Note that the definitions \rn{resecc} differ from the complex eccentricities of \citet{Lithwick2012} with the inclusion of the 
factor $\lambda_n$ (see Sect.~\ref{LXW});
this allows for the extension of their analysis for circulating systems to resonant and non-resonant librating systems.

As outlined below, the eccentricity variables $z_1$ and $z_2$ are then replaced by linear combinations
of these two quantities such that one governs the forced (or resonant) component of the interaction and the other 
governs the free (or secular) component which is not visible to the TTVs at first order.
Steps in the M1 formulation are as follows:

\begin{enumerate}
\item[(i)]
{\it Reduced disturbing function}.

The coplanar disturbing function expanded to first-order in eccentricity and reduced to the two slowly-varying
first-order resonant harmonics is
\bea
{\cal R}
&=&
m_{\rm o} a_{\rm o}^2\nu_{\rm o}^2\,[\mv_1(\alpha)\,e_{\rm i}\cos\phi_1
+\mv_2(\alpha)\,e_{\rm o}\cos\phi_2]\nextt
&=&
\ff{1}{2}m_{\rm o} a_{\rm o}^2\nu_{\rm o}^2\,[\mv_1\,z_1+\mv_2\,z_2]+CC,
\label{calR}
\eea
where $CC$ denotes the complex conjugate of the preceding expression, $\nu_{\rm o}$ is the outer mean motion,
$\alpha=a_{\rm i}/a_{\rm o}\simeq[(n+1)/n]^{-2/3}$ is the ratio of semimajor axes at exact commensurability, and
$\mv_1(\alpha)<0$ and $\mv_2(\alpha)>0$ are constants related to Laplace coefficients such that $\mv_1=-1.190$ with 
$\mv_2=0.428$ for the 2:1 commensurability. 
Importantly for the following analysis, notice the linear combination $\mv_1z_1+\mv_2z_2$ in \rn{calR}.
\vspace{0.3cm}
\item[(ii)]
{\it Evolution equations}.

From Lagrange's planetary equations and \rn{calR}, the rates of change of $z_1$ and $z_2$ are
\be
\dot z_1=i\overline\nu_{\rm i}\,\overline m_{\rm o}\,\alpha\mv_1+i(\nu_S+\dot{\tilde\theta})\,z_1
\hand
\dot z_2=i\overline\nu_{\rm o}\,\overline m_{\rm i}\,\mv_2+i(\nu_S+\dot{\tilde\theta})\,z_2,
\ee
where $\overline\nu_{\rm i}$ and $\overline\nu_{\rm o}$ are the ``mean-mean motions''.
Note that the nonlinear terms $i(\nu_S+\dot{\tilde\theta})\,z_1$ and $i(\nu_S+\dot{\tilde\theta})\,z_2$
extend the formulation of \citet{Lithwick2012} to all systems close to a first-order commensurability,
be they librating or circulating, resonant or non-resonant.
The equilibrium values of $z_1$ and $z_2$ corresponding to $\dot{\tilde\theta}=0$ (their primary fixed-point values) are then
\be
z_1^{(+)}=-\frac{\overline m_{\rm o}\alpha^{-1/2}\mv_1}{n\Delta\sigma}
\hand
z_2^{(+)}=-\frac{\overline m_{\rm i}\mv_2}{n\Delta\sigma},
\label{zplus}
\ee
so that the fixed-point eccentricities and resonance angles are
\be
\ei^{(+)}=\frac{\overline m_{\rm o}\alpha^{-1/2}|\mv_1|}{n\Delta\sigma},
\hspace{0.5cm}
\eo^{(+)}=\frac{\overline m_{\rm i}\mv_2}{n\Delta\sigma},
\hspace{0.5cm}
\phi_1^{(+)}=0
\hand
\phi_2^{(+)}=\pi.
\label{eplus}
\ee
Here $\Delta\sigma$ is the distance of the primary fixed point to exact commensurability
and is such that the superfrequency at the primary fixed point is $\nu_S^{(+)}=n\nuo^{(+)}\Delta\sigma$,
where $\nuo^{(+)}$ is the outer mean motion at this point. This quantity is determined uniquely from
the masses and elements at the reference time (M1), and can be measured if the super and resonant components
of the TTV signal can be accurately characterized \citep{Almenara2025}.
Note from \rn{phi} that to linear order, $\varpi_{\rm i}-\varpi_{\rm o}=\pi$ at the primary fixed point.
While this study has no direct use for
the equation governing the rate of change of $\dot{\tilde\theta}$, we note that $\ddot{\tilde\theta}$ 
is proportional to the linear combination $\mv_1z_1+\mv_2z_2$ (but see \rn{ODEs} below).
\vspace{0.3cm}
\item[(iii)]
{\it Rotation transformation}.

The reduced disturbing function \rn{calR} governs the interaction between the orbits to first order in eccentricity
via the linear combination $\mv_1z_1+\mv_2z_2$. It therefore makes sense to define new eccentricity variables
such that one is proportional to this quantity, while the other is an independent linear combination of $z_1$ and $z_2$
which remains constant during the evolution. 
Such a transformation was discovered by \citet{Sessin1984} using a Hamiltonian approach,
various versions of which have been employed more recently \citep[e.g.,][]{Batygin2015,Nesvorny2016,Petit2017} and each with their own scaling. 
M1 introduces the following scaling which simplifies 
the problem considerably. Defining the new variables
\be
u=-3\left(\frac{\sigma}{\Delta\sigma}\right)^2\left(\omb+ \alpha\, \omc\right)
\left(\mv_1\,z_1+\mv_2\,z_2\right)
\label{u}
\ee
and
\be
\upsilon=-3\left(\frac{\sigma}{\Delta\sigma}\right)^2\left(\omb+ \alpha\, \omc\right)
\left(g^{-1}\mv_2\,z_1-g\mv_1\,z_2\right)
\label{v}
\ee
with $\sigma=(n+1)/n$ and
$g=(\mo/\mi)^{1/2}\alpha^{-1/4}$, the evolution equations take on the especially simple form
\be
u'-i(1+\theta')u=-iR,
\hspace{0.5cm}
\upsilon'-i(1+\theta')\upsilon=0
\hand
\theta''=-\frac{1}{2i}(u-u^*),
\label{ODEs}
\ee
for which an obvious fixed point is $(u,\upsilon,\theta')=(R,0,0)$; {\it this is the primary fixed point}.
Here the positive real quantity $R$ is the {\it resonance parameter} discussed below and
the prime is $d/dx$, where $x=\nu_S^{(+)}(t-\tref)$ is the scaled ``slow'' time.
The quantity $\theta$ is related to $\ttheta$ such that
\be
n\lambda_{\rm i}-(n+1)\lambda_{\rm o}
=x+\theta
=\lambda_n(\tref)+[1+\langle\theta'\rangle]\,x+\ttheta-\ttheta(\tref),
\ee
where $\langle\theta'\rangle$ is the average value of $\theta'$ over a complete resonant cycle
(which is close to zero when the system is in the librating state and exactly zero at the primary fixed point).
Thus in general
\be
\nu_S=[1+\langle\theta'\rangle]\nu_S^{(+)},
\hspace{0.5cm}
\Delta S=[1+\langle\theta'\rangle]\Delta\sigma
\hand
\theta(\tref)=\lambda_n(\tref),
\ee
where $\nu_S=2\pi/P_S$ is the superfrequency and $\Delta S$ is the average distance
from exact commensurability.

\item[(iv)]
{\it Integrals of the motion.}

The system \rn{ODEs} admits the three integrals
\be
\ff{1}{2}(1+\theta')^2-\ff{1}{2}(u+u^*)=c_E,
\hspace{0.5cm}
-2R\theta'+uu^*=c_R=R^2
\hand
\upsilon\upsilon^*=c_F,
\label{integrals}
\ee
where $c_E$, $c_R$ and $c_F$ are constants which depend on the initial conditions, with $c_E$ increasing away from
$c_E^{({\rm min})}=\ff{1}{2}-R$, its
minimum value at the primary fixed point.
The first integral is related to the total energy, while the second and third
are such that the quantity $-2R\theta'+uu^*+\upsilon\upsilon^*$ is proportional to the
angular momentum deficit (AMD, defined relative to the primary fixed point; M1), 
allowing us the freedom to fix one of $c_R$ and $c_F$ (since it is their sum
which has physical meaning, and because $\upsilon\upsilon^*$ is constant).
In fact since the process of capture onto the primary fixed point and any subsequent (adiabatic) evolution away from it
tends to preserve the value of the second integral, it makes sense to choose $c_R=R^2$, its value at the primary 
fixed point. Then different level curves in phase-space diagrams such as in Fig.~\ref{figure:phase} correspond to 
the same value of $c_R$ but different values of $c_E$.

\hspace{0.5cm}
The first two integrals in \rn{integrals}
and the equation for $\theta''$ in \rn{ODEs} can be combined to give a nonlinear differential equation
for $\theta'$. This can be solved in terms of elliptic functions, whose period is the resonant period $P_R$
(see \citet{Nesvorny2016} who solve a related equation which arises from an alternative formulation of the problem).

\vspace{0.3cm}
\item[(v)]
{\it Resonance parameter $R$}

The resonance parameter is the value of $u$ at the primary fixed point so that from \rn{u}, \rn{zplus} and \rn{eplus},
\be
R=3\left(\frac{\sigma}{\Delta\sigma}\right)^2\left(\omb+ \alpha\, \omc\right)
(|\mv_1|\,\ei^{(+)}+\mv_2\,\eo^{(+)}).
\ee
As we show in point (vi), its value determines the resonant (or otherwise) nature of the phase space
in which a system resides.
The masses and the values of the elements at $t=\tref$ can be used to determine unique values for $\Delta\sigma$
and $R$ as well as $\theta'(\tref)$, independent of whether the system is in the librating or circulating state.
As long as nonlinear effects are unimportant, $\Delta\sigma$ and $R$ remain constant as the system evolves.

\hspace{0.5cm}
In addition, linear stability analysis reveals a relationship between $R$ and the resonant and superperiod such that
\be
R=(P_S/P_R)^2-1,
\label{PSR}
\ee
which is valid for systems close to the primary fixed point.

\vspace{0.3cm}
\item[(vi)]
{\it The primary, secondary and hyperbolic fixed points.}

As well as the primary fixed point which exists for all values of $R$, there are two other fixed points 
which exist only for $R\ge 8$ as the following shows. 
Note that $\upsilon=0$ for all three.
Since $\theta''\propto u-u^*=0$ at all fixed points, they must correspond to real values of $u$.
From the first integral in \rn{integrals} we can write
$\theta'=(u_0^2-R^2)/2R$ at the fixed points, where $u_0$ is real. Substituting this into the first equation
in \rn{ODEs} and putting $u'=0$ then gives
\be
u_0^3-R(R-2)u_0-2R^2=(u_0-R)(u_0^2+Ru_0+2R)=0,
\label{roots}
\ee
recovering the primary fixed point $u_0=u_0^{(+)}=R$ and providing simple expressions for the stable
{\it secondary fixed point}
\be
u_0^{(-)}=-\ff{1}{2}R(1-\sqrt{1-8/R})
\label{sfp}
\ee
and the unstable {\it hyperbolic fixed point}
\be
u_0^{(h)}=-\ff{1}{2}R(1+\sqrt{1-8/R}),
\label{hps}
\ee
both of which are real only when $R\ge 8$ (note that $\Delta\sigma$ and hence $R$ is positive because resonance capture
always occurs long of exact commensurability initially). Linear stability
analysis confirms the elliptic and hyperbolic nature of $u_0^{(-)}$ and $u_0^{(h)}$ respectively.
The existence of a hyperbolic fixed point renders the phase space formally resonant; hence a system is resonant
if $R\ge 8$.
The aqua, orange and blue stars in the left-hand panels of Fig.~\ref{figure:phase} correspond to the
primary, secondary and hyperbolic fixed points respectively.

\vspace{0.3cm}
\item[(vii)]
{\it Transit timing variations.}

In general, three distinct harmonics and their overtones and linear combinations\footnote{Note that 
linear combinations of non-commensurate frequencies appear
as a result of {\it nonlinear mixing} of the harmonics, introducing new frequencies \citep{Verbiest2015}. 
This effect becomes measurable in the TTV signal when the super and resonant frequencies are similar, resulting in 
the secular variation observed in systems such as {\bf Kepler-88} \citep{Nesvorny2013} and {\bf TOI-1130} \citep{Borsato2024} 
(although the effect is not identified by those authors).
It is distinct from {\it beating} which involves the linear superposition of two or more harmonics
and results in amplitude modulation (but introduces no new frequencies).}
contribute to the TTV signal: the super harmonic,
the resonant harmonic and the chopping (synodic) harmonic, whose associated periods are (in general) such that
$P_S>P_R>P_{\rm chop}$. 
Note that the planet-to-star masses ratios can be measured independent of any degeneracy in the eccentricities
and apsidal angles if at least two of these harmonics are present in the signal (see \citealt{Almenara2025} for 
a first-principles discussion of how this works).
The following discussion pertains to the super and resonant harmonics,
for which at first order there is no nonlinear interference. 

\hspace{0.5cm}
At first order the super component of the TTV signal is due to variation of the quantities $\ei\cos\varpi_{\rm i}$
and $\eo\cos\varpi_{\rm o}$ for the inner and outer orbits respectively.
The signals are anti-phased with corresponding amplitudes 
\be
A_S^{({\rm i})}=\left(\frac{\Pei}{\pi}\right)\,\ei^{(S)}
\hand
A_S^{({\rm o})}=\left(\frac{\Peo}{\pi}\right)\,\eo^{(S)}
\ee
where $\ei^{(S)}$ and $\eo^{(S)}$ are given by \rn{eplus} with $\Delta\sigma$ replaced by $|\Delta S|$,
while the period of variation is 
the superperiod $P_S$
due to the combined variation of $\ei$ and $\cos\varpi_{\rm i}$  (and similarly for the outer orbit). 
Note that for systems residing at the primary fixed point at which the eccentricities and orbital periods are constant
and $\varpi_{\rm i}-\varpi_{\rm o}=\pi$,
$\dot\varpi_{\rm i}=\dot\varpi_{\rm o}=\nu_S$ so that the TTVs are entirely due to apsidal motion.
Note also that this component is present in the signal for all configurations.

\hspace{0.5cm}
The resonant component of the TTV signal is due to energy and angular momentum exchange between the orbits 
(and hence changes in the orbital periods and eccentricities), and manifests itself through the variation of 
$\tilde\theta(t)\simeq\Theta_R\cos[\nu_R(t-\tref)+\beta_R]$ with $\nu_R=2\pi/P_R$ and $\beta_R$ the phase
at $t=\tref$.
The signals are also anti-phased with corresponding amplitudes 
\be
A_R^{({\rm i})}=\left(\frac{\Pei}{2n\pi}\right)\left(\frac{\alpha\mo}{\mi+\alpha\mo}\right)\Theta_R
\hand
A_R^{({\rm o})}=\left(\frac{\Peo}{2(n+1)\pi}\right)\left(\frac{\mi}{\mi+\alpha\mo}\right)\Theta_R,
\label{ARio}
\ee
while the period of variation is the resonant period $P_R$.
Since $\Theta_R=0$ at each of the fixed-point states, the resonant component is completely absent from the TTV signal
in such a state.
Finally notice that 
\be
A_S^{({\rm i})}/A_S^{({\rm o})}=\ei^{(S)}/\eo^{(S)}
=\alpha(|\mv_1|/\mv_2)\,(\mo/\mi)
\hand
A_R^{({\rm i})}/A_R^{({\rm o})}=\alpha\,(\mo/\mi).
\ee

\vspace{0.3cm}
\item[(viii)]
{\it Forced and free eccentricity.}

The inverse transformation of the system \rn{u} and \rn{v} 
provides natural definitions for forced and free
eccentricity which are different to those of \citet{Lithwick2012} (see Sect.~\ref{LXW} for a comparison). 
In particular the new definitions distinguish between the parts of the inner and outer eccentricity vectors
which are forced (and in general vary)
due to the presence of the companion, and the parts which are independent of its presence and 
are therefore constant (at first order).
This is achieved by writing
\bea
z_1
&=&
\left[e_b^{(+)}u/R\right]+\left[-g\,e_c^{(+)}\upsilon/R\right]\nextt
&\equiv&
z_1^{({\rm forced})}+z_1^{({\rm free})}
\label{z1ff}
\eea
and
\bea
z_2
&=&
\left[-e_c^{(+)}u/R\right]+\left[-g^{-1}e_b^{(+)}\upsilon/R\right]\nextt
&\equiv&
z_2^{({\rm forced})}+z_2^{({\rm free})},
\label{z2ff}
\eea
with the square brackets indicating the definitions and $z_1$ and $z_2$ given by \rn{resecc}.
From the integrals \rn{integrals} the forced eccentricities are then
\be
\ei^{({\rm forced})}
=
\left[z_1^{({\rm forced})}{z_1^{({\rm forced})}}^*\right]^{1/2}
=
\sqrt{1+2\theta'/R}\,\,\ei^{(+)}
\hand
\eo^{({\rm forced})}=\sqrt{1+2\theta'/R}\,\,\eo^{(+)},
\label{ebforced}
\ee
with their oscillations {\it in phase} (for both the librating or circulating states) and entering via 
the $\tilde\theta'$ component of $\theta'$,
while the constancy of $\upsilon\upsilon^*$ results in constant free eccentricities such that
\be
\ei^{({\rm free})}
=
\left[z_1^{({\rm free})}{z_1^{({\rm free})}}^*\right]^{1/2}
=
\left[g\,|\upsilon(\tref)|/R\right]\,\eo^{(+)}
\hand
\eo^{({\rm free})}=\left[g^{-1}|\upsilon(\tref)|/R\right]\,\ei^{(+)}.
\label{ebfree}
\ee
Defining the associated apsidal angles $\varpi_{\rm i}^{({\rm forced})}$ and $\varpi_{\rm i}^{({\rm free})}$ such that
\be
z_1^{({\rm forced})}=\ei^{({\rm forced})}{\rm e}^{i\,[n\lambda_{\rm i}-(n+1)\lambda_{\rm o}+\varpi_{\rm i}^{({\rm forced})}]}
\hand
z_1^{({\rm free})}=\ei^{({\rm free})}{\rm e}^{i\,[n\lambda_{\rm i}-(n+1)\lambda_{\rm o}+\varpi_{\rm i}^{({\rm free})}]}
\ee
and similarly for $\varpi_{\rm o}^{({\rm forced})}$ and $\varpi_{\rm o}^{({\rm free})}$ via 
$z_2^{({\rm forced})}$ and $z_2^{({\rm free})}$,
then writing the plane polar representations of $u$ and $\upsilon$ as
\be
u=|u|\,{\rm e}^{i\psi} 
\hand
\upsilon=|\upsilon|\,{\rm e}^{i\varphi}
\label{uandv}
\ee
where
\be
\psi=n\lambda_{\rm i}-(n+1)\lambda_{\rm o}+\varpi_\psi
\hand
\varphi=n\lambda_{\rm i}-(n+1)\lambda_{\rm o}+\varpi_\varphi,
\label{psi}
\ee
we have from \rn{z1ff} and \rn{z2ff} that
\be
\varpi_{\rm i}^{({\rm forced})}=\varpi_\psi=\varpi_{\rm o}^{({\rm forced})}+\pi
\hand
\varpi_{\rm o}^{({\rm free})}=\varpi_\varphi(T_0)+\pi=\varpi_{\rm i}^{({\rm free})}
\equiv \varpi^{({\rm free})}={\rm constant},
\label{wpsi}
\ee
where the constancy of the free apsidal angle can be demonstrated from the analytic solution for $\upsilon$ (M1).
Thus the forced components of the eccentricity vectors are anti-aligned, while the free components are aligned
and stationary (at first order). Moreover we have that
\be
\ei^{({\rm forced})}/\eo^{({\rm forced})}=\ei^{(+)}/\eo^{(+)}=\alpha^{-1/2}(\mo/\mi)|\mv_1|/\mv_2
\hand
\ei^{({\rm free})}/\eo^{({\rm free})}=\mv_2/|\mv_1|,
\ee
so that perhaps counter-intuitively, the free components, which indicate the degree to which
the system is relaxed, are not independent.
Note that the ratios $\ei^{({\rm free})}/\ei^{({\rm forced})}$ and $\eo^{({\rm free})}/\eo^{({\rm forced})}$
determine the librating and circulating behaviour of the resonance angles (M1; see also the discussion
of supercycles below).
\vspace{0.3cm}
\item[(ix)]
{\it Relaxed states.} 

A fully relaxed nonlinear mechanical system is one for which the rate of change of total energy is zero in 
the presence of dissipation \citep{Jordan2007}. This state may be completely stationary, or may be a limit-cycle
\citep[the ``overstable librations'' of][]{Goldreich2014}.
For a (Newtonian) migrating planetary system for which only length ratios (and not abolute
lengths) govern the dynamics, this is true in the co-moving frame, and requires that accelerations associated with
dissipation be of the form \rn{diss} (or similar) in order for eccentricity and period-ratio excitation to take place.

\hspace{0.5cm}
In the linear theory, all fully relaxed two-planet systems have zero free eccentricity so
that a useful measure of the degree of relaxation is the ratio 
\be
r_{\rm AMD}\equiv\upsilon\upsilon^*/(R^2+\upsilon\upsilon^*)
\ee
(see Eq.~\rn{integrals}). Values of $r_{\rm AMD}$ are listed below for various well-characterized systems.
Relaxed systems are in one of the following states:
\ben
\item[(A)]
The stationary {\it primary fixed point} at which $u=R$, $\upsilon=0$, with
the eccentricities and resonance angles given by \rn{eplus},
and the distance, $\Delta S$, from exact commensurability positive and equal to $\Delta\sigma$.
This state may be resonant or non-resonant. As discussed above:
\begin{itemize}
\item[$\star$]
{\bf GJ~876}c,b is in this state when higher-order effects are taken into account.
\end{itemize}
\item[(B)]
The {\it primary limit cycle} state at which $|u|=R\sqrt{1+2\theta'/R}$, $\theta'\ne 0$, and $\upsilon=0$,
with the eccentricities oscillating in phase with period $P_R$, the resonance angles $\phi_1$ and $\phi_2$
librating in phase with period $P_R$ around zero and $\pi$ respectively if the
system is in the librating state, and
$\varpi_{\rm i}-\varpi_{\rm o}=\pi$ exactly for both librating and circulating systems. The system may be 
resonant or non-resonant,
although simulations which use simple disk models suggest that
resonant systems which reach the separatrix during the capture process tend to move on to the
next commensurability.\footnote{{\bf Kepler-9} is a curious counter-example consisting of a pair of $0.1 M_J$ planets 
\citep{Borsato2019} in the circulating state just outside the 2:1 separatrix in a severely-resonant phase space (M1).
Note that $\upsilon\ne 0$ for this system.}
Examples near this state include:
\begin{itemize}
\item[$\star$]
The {\cyan non-resonant librating} system {\bf TOI-1408}c,b \citep{Korth2024} with parameters
$\Peo/\Pei=2.04$, $m_*=1.3 M_\odot$, $\mi=0.02 M_J$, $\mo=1.9 M_J$, 
$R=3.4$, $\ei=0.14$, $\eo=0.002$, $\ei^{({\rm free})}=0.0008$, $\eo^{({\rm free})}=0.002$ and $r_{\rm AMD}=0.04$;

\item[$\star$]
The {\red resonant librating} system {\bf TOI-216}b,c \citep{Nesvorny2022} with parameters
$\Peo/\Pei=2.02$, $m_*=0.77 M_\odot$, $\mi=0.06 M_J$, $\mo=0.6 M_J$, 
$R=21.1$, $\ei=0.16$, $\eo=0.005$, $\ei^{({\rm free})}=0.002$, $\eo^{({\rm free})}=0.006$ and $r_{\rm AMD}=0.02$.

\item[$\star$]
{\bf TOI-4504}d,c {\it appears to be in a higher-order limit-cycle state}, with the $\upsilon$-component

librating around its non-zero equilibrium value rather than fixed at zero.
By simulating the capture process for a system with the same masses and with disk acceleration 
given by \rn{diss},\footnote{We also included a disk edge to ensure 
convergent migration given the planet-planet mass ratio is only 1.3.} 
we verified that the phase space behaviour seen in Fig.~\ref{figure:phase} is indeed a limit-cycle state.
\end{itemize}
\item[(C)]
The stationary {\it secondary fixed point} (which exists only for $R\ge 8$) at which $u=u_0^{(-)}$,
$\Delta S=\Delta\sigma^{(-)}\equiv u_0^{(h)}\Delta\sigma<0$ (where $u_0^{(h)}$ is given in terms of $R$ by \rn{hps}),
the eccentricities are given by \rn{eplus} but with $\Delta\sigma$ replaced by $\Delta\sigma^{(-)}$,
and the resonance angles are $\phi_1=\pi$ and $\phi_2=0$. 
An example which is very close to this state is:

\begin{itemize}
\item[$\star$]
{\bf Kepler-18}c,d \citep{Cochran2011} (2:1; {\red resonant librating}, $\Delta S=-0.06$, $r_{\rm AMD}=5\times 10^{-5}$).      
\end{itemize}
\item[(D)]
The {\it secondary limit-cycle} state is similar to the primary limit-cycle state but with $|u|=|u_0^{(-)}|\sqrt{1+2\theta'/R}$.
An example of a system close to this state is:
\begin{itemize}
\item[$\star$] 
{\bf TOI-7510}b,c \citep{Almenara2025b} (2:1; {\red resonant circulating}, $\Delta S=-0.04$, $r_{\rm AMD}=0.06$).      
\end{itemize}
Note that both Kepler-18c,d and TOI-7510b,c participate in near-Laplace-like configurations (4:2:1).
\een

\vspace{0.3cm}
\item[(x)]
{\it Supercycles.}

Supercycles are completely relaxed in $u$-space but not in $\upsilon$-space,
suggesting that the formation damping timescales are different for each mode, and tend to have high values of $r_{\rm AMD}$
which distinguishes them from limit cycles for which $r_{\rm AMD}=0$ at first order.
As a result the resonant component of the TTVs is completely absent, and
since at first order the free eccentricity and apsidal angle are constant, 
TTVs can only distinguish this state when (higher-order) nonlinear interference between the resonant and secular modes
is significant enough to cause the free quantities to vary.
The primary and secondary supercycle states each have four distinct sub-states:
\ben
\item[(A)]
The primary supercycle state is defined by $u=u_0^{(+)}=R$, $\upsilon\ne 0$, that is, the resonant ($u$) component is 
identical to the primary fixed point state but the free eccentricity is non-zero. As a result the period ratio is constant
(as are the mean motions), while the eccentricities are anti-phased and vary with period $P_S$.
The average value of the inner eccentricity is $\ei^{(+)}$ if $\ei^{({\rm free})}<\ei^{(+)}$ with amplitude 
$\ei^{({\rm free})}$ and $\phi_1$ librating, while the average value 
is $\ei^{({\rm free})}$ if $\ei^{({\rm free})}>\ei^{(+)}$ with amplitude 
$\ei^{(+)}$ and $\phi_1$ circulating. The situation is similar for the outer eccentricity and $\phi_2$,
resulting in the four possible states $(l,l), (l,c), (c,l)$ and $(c,c)$, where $l$ and $c$ stand for the libration or circulation
of $(\phi_1,\phi_2)$.
Note that not all four are possible for a given value of $\mo/\mi$.

\hspace{0.5cm}
Using published parameters (or the whole posterior for K2-19), systems which are close to this state include:
\begin{itemize}
\item[$\star$] 
{\bf Kepler-36}b,c \citep{Vissapragada2020} (7:6, $R=2.5$, {\cyan librating non-resonant}, $(c,c)$, $r_{\rm AMD}=0.99$);
\item[$\star$] 
{\bf K2-19}b,c \citep{Almenara2025} (3:2, $R=15.5$, {\red librating resonant}, $(l,l), (c,l), (c,c)$, $r_{\rm AMD}=0.85$);
\item[$\star$] 
{\bf TOI-1130}b,c \citep{Borsato2024} (2:1, $R=0.23$, librating non-resonant, $(l,c)$, $r_{\rm AMD}=0.95$); 
\item[$\star$] 
{\bf The Pluto-Neptune system} \citep{Malhotra1993,Murray1999} (3:2, $R=3481$, {\red librating severely resonant}, $(l,l)$, $r_{\rm AMD}=0.89$). 
\end{itemize}
\item[(B)]
The secondary supercycle state is defined by $u=u_0^{(-)}$, $\upsilon\ne 0$, that is, the resonant component is 
identical to the secondary fixed point state but the free eccentricity is non-zero. 
A system which is very close to this state is:
\begin{itemize}
\item[$\star$] 
{\bf K2-266}d,e  \citep{Jiang2025} (4:3; $\Delta S=-0.007$, $R=10.5$, {\red librating resonant}, $(c,c)$, $r_{\rm AMD}=0.83$).
\end{itemize}
\een

\end{enumerate}

\subsection{Second-order theory}\label{second}

The purpose of the following is to illustrate how nonlinear mixing of the resonant and secular modes is introduced
at second order, and omits contributions 
from the variation of the mean longitude at epoch and the term involving $\partial{\cal R}/\partial\lambda$
in Lagrange's planetary equation for the rate of change of the eccentricity.
Note that the latter terms are second-order in $u$ and $\upsilon$ in \rn{uv2nd}.

To second-order in eccentricity, three
additional slowly-varying harmonics involving $2\phi_1$, $\phi_1+\phi_2$ and $2\phi_2$ are introduced and
the reduced disturbing function \rn{calR} becomes
\be
{\cal R}=
\ff{1}{2}m_{\rm o} a_{\rm o}^2\nu_{\rm o}^2\,[\mv_1\,z_1+\mv_2\,z_2
\,\,{\blue +\,\,w_1\,z_1^2+2w_{12}\,z_1z_2+w_2\,z_2^2}]+CC,
\ee
where $w_1(\alpha)>0$, $w_{12}(\alpha)<0$ and $w_2(\alpha)>0$ are constants related to Laplace coefficients such that
$w_1=1.696$, $2w_{12}=-4.967$ and $w_2=3.594$
for the 2:1 commensurability. 
Substituting the inverse rotation transformation \rn{z1ff} and \rn{z2ff} into the resulting equations for the
rates of change of $z_1$, $z_2$ and $\dot\theta$ (and scaling the time)
then yields evolution equations for $u$, $\upsilon$ and $\theta'$ such that
\be
u'-i(1+\theta')u\,\,{\blue-\,\,i(D_{11}u^*+D_{12}\upsilon^*)}=-iR,
\hspace{0.5cm}
\upsilon'-i(1+\theta')\upsilon\,\,{\blue-\,\,i(D_{12}u^*+D_{22}\upsilon^*)}=0,
\label{uv2nd}
\ee
and
\be
\theta''=-\frac{1}{2i}\left[u\,\,{\blue-\,\,(D_{11}u^2+2D_{12}u\,\upsilon+D_{22}\upsilon^2)/R}\right]+CC,
\label{ODEs2}
\ee
with integrals
\be
\frac{1}{2}(1+\theta')^2-\frac{1}{2}(u+u^*)
\,\,{\blue +\,\frac{1}{4R}\left[D_{11}(u^2+{u^*}^2)+2D_{12}(u\,\upsilon+u^*\upsilon^*)+D_{22}(\upsilon^2+{\upsilon^*}^2)\right]}
=c_E^{(2)}
\label{cE2}
\ee
and
\be
-2R\theta'+{\blue uu^*+\upsilon\upsilon^*}=c_J,
\ee
where $c_E^{(2)}$ and $c_J$ are constants of the motion.
Here the terms in blue introduce mixing of the resonant and secular modes,
while the $D_{ij}\propto\Delta\sigma^{-1}$ 
depend on the $\mv$'s and $w$'s as well as the planet-to-star mass ratios {\it and} the
planet-to-planet mass ratios.
For example, in the limit of large $\mo/\mi$ (and therefore large $g=\alpha^{-1/4}(\mo/\mi)^{1/2}$) 
as is the case for TOI-1130 and TOI-1408 (see Table~\ref{table:second}), we have
\be
\lim_{g^{-1}\rightarrow 0}D_{11}=e_b^{(+)}d_{11}^{\,(g\uparrow)},
\hspace{0.5cm}
\lim_{g^{-1}\rightarrow 0}D_{12}=g^{-1}e_b^{(+)}d_{12}^{\,(g\uparrow)}
\hand
\lim_{g^{-1}\rightarrow 0}D_{22}={\cal O}(g^{-2}),
\ee
where $d_{11}^{\,(g\uparrow)}=2w_1/|\mv_1|$ is 2.85 for 2:1 and 5.27 for 3:2,
while $d_{12}^{\,(g\uparrow)}=2(w_{12}/\mv_1-\mv_2w_1/\mv_1^2)$ is 3.15 for 2:1 and 0.11 for 3:2.
Similarly for small $\mo/\mi$ (small $g$),
\be
\lim_{g\rightarrow 0}D_{11}=e_c^{(+)}d_{11}^{\,(g\downarrow)},
\hspace{0.5cm}
\lim_{g\rightarrow 0}D_{12}=g\,e_c^{(+)}d_{12}^{\,(g\downarrow)}
\hand
\lim_{g\rightarrow 0}D_{22}={\cal O}(g^2),
\ee
where $d_{11}^{\,(g\downarrow)}=2w_2/\mv_2$ is 16.78 for 2:1 and 6.65 for 3:2,
while $d_{12}^{\,(g\downarrow)}=2(w_{12}/\mv_2-\mv_1w_2/\mv_2^2)$ is 35.03 for 2:1 and 0.07 for 3:2.

Table~\ref{table:second} 
\begin{table*}
\tiny
  \setlength{\tabcolsep}{5pt} 
\renewcommand{\arraystretch}{0.95}
\caption{Second-order parameters for various systems, with the magnitude of $D_{12}$ indicative of
the strength of the nonlinear interaction}             
\label{table:second}      
\centering                          
\begin{tabular}{l c c c c c c c r}        
\hline\\
System & $n+1:n$ & $R$ & $\Delta\sigma$ & $m_*,\,\mi,\,\mo$ &
$D_{11}$ & $D_{12}$ & $D_{22}$ &  $\upsilon^{(+)}$ 
 \\    
\hline                        
&\\
TOI-4504d,c & 2:1 & 50 & 0.02 & 0.8, 2.0, 2.5  &  1.1 & {\bf 0.7} & 0.5 &  $-14.2$  \\
&\\
GJ 876c,b & 2:1 & 13.6 & 0.05 & 0.3, 0.9, 2.8  & 0.9 & {\bf 0.4} & 0.2 &  $-2.9$  \\
 &  \\
TOI-1130b,c & 2:1 & 0.2 & 0.05 & 0.8, 0.06, 1.1 & 0.1 & {\bf 0.03} & 0.007 &  $-0.006$  \\
 &  \\
TOI-1408c,b & 2:1 & 3.4 & 0.02 & 2.3, 0.02, 1.9  & 0.3 & {\bf 0.03} & 0.004 &  $-0.09$  \\
 &  \\
K2-19b,c & 3:2 & 15.5 & 0.003 & 0.9, 0.09, 0.03 & 0.003 & ${\bf -0.002}\phantom{-}$ & 0.001 &  0.03 \\
 &  \\
Neptune-Pluto & 3:2 & 3481 & $0.0003$ & 1.0, 0.05, $7\times 10^{-6}$ 
& 1.7 & ${\bf 0.0002}$ & $-6\times 10^{-7}$ &  $-0.3$ \\
 &  \\
\hline                                   
\end{tabular}
\tablefoot{$R$ and $\Delta\sigma$ from first-order theory. Mass units: star $M_\odot$, planets $M_J$. 
}
\end{table*}
lists values of $D_{ij}$ for TOI-4504 and GJ~876, as well as the non-resonant librating systems
TOI-1130 \citep{Borsato2024} and TOI-1408 \citep{Korth2024}, both of which are near the 2:1 commensurability and 
harbour a giant and a Neptune-like planet. 
In addition to excellent photometric precision, nonlinear mixing allows for 
the full eccentricities and apsidal angles to be well characterized for each of these systems, thanks especially
to the magnitude of $D_{12}$ (see Eqs.~\rn{uv2nd} to \rn{cE2} and boldface values in Table~\ref{table:second}).
In particular, the contribution to $\theta''$ from the term proportional to $D_{12}u\upsilon$ in \rn{ODEs2} 
modifies $\Theta$ and hence the TTV resonant amplitudes \rn{ARio}, distinguishing between systems
with different eccentricities and apsidal angles.

In contrast, $D_{12}$ tends to be small for systems near the $n+1:n$ commensurability for $n\ge 2$,
as is the case for K2-19 which is also listed in Table~\ref{table:second}. As a result, the posterior projections
of the eccentricities and apsidal angles tend to be highly degenerate for such systems 
(see Fig.~6 of \citealt{Almenara2025} for K2-19).

Also listed in Table~\ref{table:second} are second-order estimates for the fixed-point values of 
\be
\upsilon=\upsilon^{(+)}=-RD_{12}/|D|,
\ee
obtained by putting $u'=\upsilon'=\theta'=0$ in \rn{uv2nd}. Here $|D|=(1+D_{11})(1+D_{22})-D_{12}^2$,
These are around half the true values for both TOI-4504 and GJ~876 (see Fig.~\ref{figure:phase}), 
which we attribute to the neglect of higher-order terms, but are nonetheless several orders of magnitude higher
than for the other systems listed.

\subsection{Forced and free eccentricity in the \citet{Lithwick2012} formulation}\label{LXW}

\citet{Lithwick2012} (hereafter LXW) introduce the concepts of forced and free eccentricity for near-resonant systems,\footnote{Note that 
these are distinct from the definition given in \citet{Murray1999} for secular systems.}
defining the inner and outer ``complex eccentricities'' (using our notation as well as $\zeta_1$ and $\zeta_2$
for their $z$ and $z'$ to avoid confusion) as $\zeta_1=\ei\,{\rm e}^{i\varpi_{\rm i}}$
and $\zeta_2=\eo\,{\rm e}^{i\varpi_{\rm o}}$. Notice these differ from our definitions of $z_1$ and $z_2$ in \rn{resecc}
which involve the additional factor ${\rm e}^{i\lambda_n}$ (note also that their 
$\lambda^j=-(j-1)\lambda_{\rm i}+j\lambda_{\rm o}$ is our $-\lambda_n=-n\lambda_{\rm i}+(n+1)\lambda_{\rm o}$).
The authors then decompose the complex eccentricities into ``forced'' and ``free'' components such that
\be
\zeta_1=\zeta_1^{({\rm forced})}+\zeta_1^{({\rm free})}
\hand
\zeta_2=\zeta_2^{({\rm forced})}+\zeta_2^{({\rm free})},
\ee
where
\be
\zeta_1^{({\rm forced})}=s\,\ei^{(S)}\,{\rm e}^{-i\lambda_n}=-\zeta_2^{({\rm forced})}
\ee
with $s={\rm sgn}(\Delta S)$, and
$\zeta_1^{({\rm free})}$ and $\zeta_2^{({\rm free})}$ are integration constants (which come from
integrating their Eq.~(A13)), and remark that ``the complex free eccentricities precess on a secular timescale''
(i.e., on a timescale much longer than the observations).
Note that the quantity $\Delta\equiv \Delta_{\rm LXW}$ in LXW is such that $(n+1)\Delta_{\rm LXW}=n\Delta S$.
Perturbations to the mean longitudes are derived effectively from conservation of angular momentum,\footnote{In fact
they use two related integrals.} after which they arrive at expressions for the inner and outer TTVs
\be
TTV_{\rm i,o}^{{\rm LXW}}=
\left(\frac{P_{\rm i,o}}{2\pi}\right)|E_{\rm i,o}^{{\rm LXW}}|
\sin[\nu_S(t-\tref)+\beta_{\rm i,o}],
\label{TTV-LXW}
\ee
where
\be
E_{\rm i}^{{\rm LXW}}=\ei^{(S)}\left[1-\frac{3}{2}\frac{1}{|\mv_1|}\frac{Z_{\rm free}^*}{\Delta_{\rm LXW}}\right]
\hand
E_{\rm o}^{{\rm LXW}}=-\eo^{(S)}\left[1-\frac{3}{2}\frac{1}{\mv_2}\frac{Z_{\rm free}^*}{\Delta_{\rm LXW}}\right],
\label{Eamp}
\ee
for which LXW define the complex quantity
\be
Z_{\rm free}=\mv_1\zeta_1^{({\rm free})}+\mv_2\zeta_2^{({\rm free})}.
\label{Zfree}
\ee
But this is the combination of eccentricities appearing in the disturbing function \rn{calR}
and in the definition of $u$ (Eq.~\rn{u}), so the sense of ``free'' in the M1 and LXW formulations is quite different.
Indeed for M1, ``free'' refers to the part of the signal which is not detectable at first order and which involves 
the linear combination appearing in \rn{v} rather than \rn{Zfree}, while for LXW this component 
does indeed affect the signal.
In fact LXW make it clear that their formulation is valid only when $|Z_{\rm free}/\Delta_{\rm LXW}|$ is very large
or very small. The first limit corresponds to circulating systems far outside the separatrix/libration boundary
where $|u|$ is approximately constant and $\langle\dot\varpi_\psi\rangle\ll \nu_S$, justifying the statement that 
the complex free eccentricities evolve secularly, while the second limit corresponds to either of the 
stable fixed points. Moreover, while librating systems and circulating systems close to the libration boundary vary
with period $P_R$, far from the libration boundary $P_R\simeq P_S$, justifying the single harmonic in \rn{TTV-LXW}
and causing the mass-eccentricity degeneracy inherent in many system which exhibit TTVs.

While the two TTV formulations are identical at the stable fixed points 
(where $\Theta_R$ and $Z_{\rm free}$ are zero and from \rn{ebforced} the forced eccentricities are the same),
they can be reconciled for systems far outside the libration boundary 
if we define
\be
Z_{\rm forced}\equiv \left(\mv_1 z_1^{({\rm forced})}+\mv_2 z_2^{({\rm forced})}\right){\rm e}^{-i\lambda_n},
\ee
where $z_1^{({\rm forced})}$ and $z_2^{({\rm forced})}$ are given in \rn{z1ff} and \rn{z2ff}
(notice that $Z_{\rm forced}$ is independent of $\lambda_n$),
take $Z_{\rm forced}$ and the mean motions to be constant,
and integrate \hbox{$\theta''=\ff{1}{2}i(u-u^*)$} twice (the third equation in \rn{ODEs}).
The TTV amplitudes are then \rn{Eamp} with $Z_{\rm free}$ replaced by $Z_{\rm forced}$.

On the other hand, the LXW formulation says nothing about the free component associated with $\upsilon$,
whose magnitude is truly constant at first-order in eccentricity, and varies secularly only at second order through 
nonlinear interference of the resonant and secular modes.

\end{appendix}
\end{document}